\documentclass[preprints,article,submit,moreauthors,pdftex]{Definitions/mdpi} 

\renewcommand{\linenumbers}{}
\usepackage{graphics,epsfig}
\usepackage{epstopdf}
\usepackage{changes}
\usepackage[toc,page]{appendix}
\usepackage[normalem]{ulem}
\usepackage{adjustbox}
\usepackage{latexsym}
\usepackage{amsmath}
\usepackage{amssymb}
\usepackage{amsfonts}
\usepackage{footnote}
\usepackage{times}
\usepackage{graphicx}
\usepackage{dcolumn}
\usepackage{amsmath}
\usepackage{array}
\usepackage{wrapfig}
\usepackage{multirow}
\usepackage{tabularx}
\usepackage{comment}
\usepackage{verbatim}

\def\ber{\begin{eqnarray}}
\def\eer{\end{eqnarray}}
\def\beq{\begin{equation}}
\def\eeq{\end{equation}}

\newcommand\vep{\varepsilon}

\firstpage{1} 
\pubvolume{1}
\issuenum{1}
\articlenumber{0}
\pubyear{2022}
\copyrightyear{2022}
\externaleditor{} 
\datereceived{} 
\dateaccepted{} 
\datepublished{} 
\hreflink{https://doi.org/} 

\Title{Deflection angle and shadow of the Reissner-Nordstr\"om
black hole with higher-order magnetic correction 
in Einstein-nonlinear-Maxwell fields}

\TitleCitation{Deflection angle and shadow of the Reissner-Nordstr\"om
BH with higher-order magnetic correction 
in ENM fields}

\Author{{Yashmitha Kumaran} 
$^{1,\dag}$, and Ali {\"O}vg{\"u}n \orcidB{ }{$^{1,\dag}$} 
}

\AuthorNames{Yashmitha Kumaran and Ali {\"O}vg{\"u}n}

\AuthorCitation{Kumaran, Y.; {\"O}vg{\"u}n, A.}

\address{%
$^{1}$ \quad {Physics Department, Eastern Mediterranean University, North Cyprus via Mersin 10, 
{Famagusta~99628},~Turkey} ; {yashmitha.kumaran@emu.edu.tr (Y.K.); }{ali.ovgun@emu.edu.tr (A.O.)} 
}
\corres{Correspondence: {ali.ovgun@emu.edu.tr}}

\firstnote{{These authors contributed equally to this work.}}

\abstract{Nonlinear electrodynamics is known as the generalizations of Maxwell electrodynamics at strong fields and presents interesting features such as curing the classical divergences present in the linear theory when coupled to general relativity. In this paper, we consider the asymptotically flat Reissner-Nordstr\"om black hole solution with higher-order magnetic correction in Einstein-nonlinear-Maxwell fields. We study the effect of the magnetic charge parameters on the black hole, viz. weak deflection angle of photons and massive particles using Gauss-bonnet theorem. Moreover, we apply the Keeton-Petters formalism to confirm our results of the weak deflection angle. Apart from vacuum, their influence in the presence of different media such as plasma and dark matter are probed as well. Finally, we examine the black hole shadow cast using the null-geodesics method and investigate its spherically in-falling thin accretion disk. Our inferences show how the magnetic charge parameter $p$ affects the other physical quantities; so, we impose some constraints on this parameter using the observations from the Event Horizon Telescope.}



\begin{document}

\section{Introduction}
Gravity is the weakest force in the universe that we live in today. While it can pull everything reachable inwards, its effect decreases with (the square of) distance until where it is no longer significant, according to the Newtonian physics. However, when the mass of the gravitating object increases, gravity starts to behave differently. With a mass high enough, collapsing on itself and concentrated at a single point, it can not only overcome the other three natural forces, but also crush the familiar laws of physics that are known to govern our universe. This object of extreme mass, infinite density, no volume and dominating gravity -- so strong that nothing that goes in comes out -- is a black hole \cite{Einstein:1916vd}. Black holes have been particularly of interest since their discovery by the Event Horizon Telescope \cite{EventHorizonTelescope:2019dse,EventHorizonTelescope:2022xnr}.

A black hole is surrounded by an accretion disk in which matter, dust and photons are stuck in unstable orbits around it \cite{Synge1966,Luminet1979}. The not-so-circular photon sphere gives rise to the phenomenon of gravitational lensing. When a massive cluster falls in the light trajectory aimed at an observer, the gravitational fields of the clusters act as a lens by deflecting the rays of light, causing distortions of the light source in its background \cite{Bozza:2002zj}. This fascinating phenomenon was particularly prominent in the first images from the James Webb Space Telescope, especially of the galaxy cluster called SMACS 0723, which was reportedly due to an astronomical quantity of matter in view on the speck of sky almost as big as a sand grain at arm-length \cite{webb}. 

Gravitational lensing can be classified into strong lensing and weak lensing; this paper is built on the latter. It majorly depends on the mass distribution of the lensing cluster. Weak lensing is a consequence of general relativity arising from minor distortions that are too small to be detected in terms of magnification, yet sufficient enough to distinguish between various mass distributions \cite{Virbhadra:1999nm,Virbhadra:2002ju,Virbhadra:1998dy,Virbhadra:2007kw,Virbhadra:2008ws,Adler:2022qtb,Hasse:2001by,Perlick:2003vg,Perlick:2018,He:2020eah}. It is known in astrophysics that distances have a dominant role in obtaining the properties of astrophysical objects. However, Virbhadra proved that just observation of relativistic images  can also say an incredibly accurate value for the upper bound to the compactness of massive dark objects \cite{Virbhadra:2022ybp} and then Virbhadra showed that there exists a distortion parameter such that the signed sum of all images of singular gravitational lensing of a source identically vanishes by testing this with images of 
Schwarzschild case (SC) lensing in weak and strong gravitational fields \cite{Virbhadra:2022iiy}.

Weak lensing utilizes the fine property of differential deflection exhibited by the bending of light to explore the structures of the cosmic deeper. In order to achieve this, the angle of deflection is calculated using the optical geometry derived from the Gauss-Bonnet theorem given by
\cite{Gibbons:2008rj}
\begin{equation}
    \int \int_D \mathcal{K} \mathrm{~d}S + \int_{\partial D} \kappa \mathrm{~d}t + \sum_i \alpha_i = 2\pi \chi(D), \label{GBT1}
\end{equation}
where, $\chi$ is the Euler characteristic of the topology, $g$ is a Riemannian metric of the manifold of the symmetric lens, $(D, \chi, g)$ represent the domain of the surface, $\mathcal{K}$ is the Gaussian curvature, $\kappa$ is the geodesic curvature, and $\alpha_i$ is the exterior angle at the $i$\textsuperscript{th} vertex. In the literature, there are various studies of this method on black holes, wormholes and other spacetimes  \cite{Werner:2012rc,Ovgun:2018fnk,Ovgun:2019wej,Ovgun:2018oxk,Li:2020dln,Li:2020wvn,Kumaran:2019qqp,Kumaran:2021rgj,kumaran3,Ovgun:2018tua,Okyay:2021nnh,Javed:2019kon,Javed:2019rrg,Javed:2019ynm,Javed:2020lsg,Javed:2019qyg,Javed:2019jag,Ishihara:2016vdc,Takizawa:2020egm,Ono:2019hkw,Ishihara:2016sfv,Ono:2017pie,Pantig:2020odu,Rayimbaev:2022hca,Pantig:2022toh,Pantig:2022ely,Pantig:2022whj,Uniyal:2022vdu,Javed:2022kzf,Jusufi:2017lsl,Javed:2021ymu,Javed:2020pyz,Ovgun:2020yuv,ElMoumni:2020wrf,Javed:2020fli,Jusufi:2017mav,Fu:2021akc,Crisnejo:2018uyn}.

The scope of this paper extends to evaluate the weak deflection angle through two different approaches: the Gibbons and Werner (GW) method \cite{Gibbons:2008rj} and the Keeton-Petters formalism \cite{Keeton:2005jd}. Moreover, we study the shadow cast of the black hole with thin-accretion disk.

The accretion disk, along with the lensing effect, creates the appearance of a shadow of the black hole. This is due to the emission region that is geometrically thick but optically thin and is accompanied by a distant, homogeneous, isotropic emission ring \cite{Jaroszynski:1997bw,Bambi:2012tg,Kruglov:2020tes}.

The shadow is essentially illustrated as the critical curve interior which separates the capture orbits that spiral into the black hole from the scattering orbits that swerve away from the black hole, i.e. entering versus exiting photon orbits. Although the size of the shadow is primarily dependant on the intrinsic parameters of the black hole and its contour is determined by the orbital instability of the light rays from the photon sphere, it merely appears to be a dark, two-dimensional disk for a distant observer illuminated by its bright, uniform surrounding \cite{Allahyari:2019jqz,Vagnozzi:2022moj,Roy:2021uye,Vagnozzi:2019apd,Khodadi:2021gbc,Khodadi:2020jij,Kumar:2018ple,Khodadi:2022pqh,Lambiase:2022ywp, Kumar:2020owy,Rahaman:2021web,Belhaj:2020rdb,Belhaj:2020okh,Belhaj:2021rae,Guo:2021bhr,Sun:2022wya,Gralla:2019xty,Ma:2020dhv,Pantig:2022ely,Ovgun:2020gjz,Pantig2020b,PANTIG2022168722,Pantig:2022toh,Ovgun:2021ttv,Okyay:2021nnh,Cimdiker:2021cpz,Pantig:2022whj,Kuang:2022xjp,Uniyal:2022vdu,Ovgun:2018tua,Herdeiro:2021lwl,Shaikh:2018lcc,Shaikh:2019fpu,Cunha:2019hzj,Cunha:2019ikd,Cunha:2018acu,Cunha:2016wzk,Vincent:2016sjq,Afrin:2021imp,Jha:2021bue,Zeng:2020dco,He:2022yse,Dokuchaev:2020wqk,Bambi:2019tjh,Meng:2022kjs,Chen:2022lct,Chen:2022nbb,Wang:2022kvg,Bronzwaer:2021lzo,Falcke:1999pj,Wei:2019pjf,Wei:2018xks,Abdolrahimi:2015rua,Adair:2020vso,Abdolrahimi:2015kma,Konoplya:2020bxa,Konoplya:2019sns,Konoplya:2019xmn,Chakhchi:2022fls,Perlick:2018iye,Perlick:2021aok,Clifton:2020xhc}.

A factor of interest that affects the radius of the shadow is the effect of magnetic charge especially since it cannot be neutralized with regular matter unlike electric charge in a conductive medium \cite{Maldacena:2020skw}. The presence of magnetic charge tends to increase the curvature of the spacetime, resulting in more photons being pulled into the black hole and hence, decreasing the radius of the shadow \cite{Guo:2021bhr}. Along with black hole spin, magnetic charge is found to create two distinct horizons, namely the inner and the outer horizons, defined by stable and unstable photon orbits  \cite{Sun:2022wya}. For a negligible charge, they are one and indistinguishable. But as the value of the charge increases towards a critical value, Sun et al. have obtained that these horizons become more prominent as the inner horizon appears from the center, with the outer horizon existing sensitive to the charge. Other studies have shown the influence of magnetic charge on the shadow for different cases of black holes \cite{Guo:2021bhr,Zhang:2021oee,Allahyari:2019jqz,Vagnozzi:2022moj,Kruglov:2020tes}.

Here, we will proceed to discuss how magnetic charges affect the shadow of a black hole starting with the black hole solution from a new model of nonlinear electrodynamics proposed by \cite{Mazharimousavi:2021uki} coupled in Einstein's gravity. The paper is organized as follows: the black hole solution is introduced in section \ref{bhs}. This is followed by calculating the weak deflection angle using the Gauss-Bonnet theorem in section \ref{wdagbt} along with determining the deflection angle and the observables using Keeton-Petters formalism. Then, the weak deflection angle for massive particles is computed in section \ref{wdajm} with the help of the Jacobi metric. Furthermore, the weak deflection angle is calculated in the presence of plasma and dark matter in section \ref{wdapdm}. With inferences and comments about shadows in section \ref{mcsbh}, we finish with concluding remarks about our results in section \ref{conc}.

\section{Brief Review of Reissner-Nordstr\"om black hole with higher-order magnetic correction in Einstein-nonlinear-Maxwell fields}
\label{bhs}
From the beginning of the universe to the black holes, singularity has been a matter of question, hindering general relativity from being unified to the other models of the universe. Researchers have been working on finding the solution for a black hole without singularities pioneered by Bardeen \cite{Bardeen,Frolov:2017dwy}.

Building on this attempt to eliminate singularity, the black hole solution presented by \cite{Mazharimousavi:2021uki}
brings nonlinear electrodynamics into play. They have provided an analytical black hole solution much like the Born-Infeld-type corrections to the linear Maxwell's theory in the weak field limit \cite{Falciano:2019fns}. The governing equations for a pure magnetic field are

\begin{equation}
    \begin{split}
    \label{ged}
         \textrm{Vector potential: } A_\phi =& \;-p \cos\theta, \\ \textrm{Spacetime tensor field: } F_{\mu\nu} =& \; \partial_\alpha A_\beta - \partial_\beta A_\alpha, \\
         \textrm{Electromagnetic invariant: } \mathcal{F} \equiv& \; \frac{1}{4} F_{\mu\nu} F^{\mu\nu}, \\
         \textrm{Lagrangian density of the new model: }\mathcal{L} =& \; -\frac{1}{\beta} \ln{\left[\cos^2\left(\sqrt{-\beta\mathcal{F}} \right) \right]} \quad , \quad \mathcal{F} > -\frac{\pi^2}{4\beta},\\
         \textrm{Action coupling }\mathcal{L} \textrm{ minimally to Einstein's gravity: }I=&\;\int \mathrm{~d}^{4} x \sqrt{-g}\left(\frac{R}{16\pi G} +\mathcal{L}\right),\\
         \textrm{Energy-momentum tensor: } T_{\mu}^{v} =& \; \frac{1}{4\pi} \left(\mathcal{L} \delta_{\mu}^{v}-\mathcal{L}_{\mathcal{F}} F_{\mu \lambda} F^{v \lambda}\right), 
    \end{split}
\end{equation}

where, $p$ is the magnetic charge construed as a magnetic monopole, $\beta$ is a dimensional constant with a dimension of $[\textrm{Length}]^4$, $g$ is the metric, $R$ is the Ricci scalar, $G$ is the Newton's gravitational constant in four-dimensional spacetime and $
\mathcal{L}_{\mathcal{F}}=\frac{\partial \mathcal{L}}{\partial \mathcal{F}}
$.

The line element of a spherically symmetric spacetime is written as
\begin{equation}
    \mathrm{~d} s^{2}=-f(r) \mathrm{~d} t^{2}+\frac{1}{f(r)} \mathrm{~d} r^{2}+r^{2}\left(\mathrm{~d} \theta^{2}+\sin ^{2} \theta \, \mathrm{~d} \phi^{2}\right).
    \label{le}
\end{equation}
Given a radial magnetic field, $B_r = p/r^2$, using the set of equations from (\ref{ged})
\begin{equation}\mathcal{F} = \frac{p^2}{2r^4},\end{equation}
and new nonlinear electrodynamics Lagrangian is proposed in \cite{Mazharimousavi:2021uki} by Mazharimousavi and Halilsoy - satisfying the Maxwell-nonlinear equations $F_{[\alpha \beta, \gamma]}=0$ and $ \partial_{\mu}\left(\sqrt{-g} \mathcal{L}_{\mathcal{F}} F^{\mu \nu}\right)=0$  - to be
\begin{equation}\mathcal{L}=-\frac{1}{\beta} \ln \left[\cosh ^{2}\left(\sqrt{\frac{\beta}{2}} \frac{p}{r^{2}}\right)\right].\end{equation}

Then the corresponding energy-momentum tensor for the nonlinear electrodynamics is calculated as follows \begin{equation}
T_{t}^{t}=-\frac{1}{4 \pi \beta} \ln \left(\cosh ^{2}\left(\sqrt{\frac{\beta}{2}} \frac{p}{r^{2}}\right)\right),
\end{equation} on the other hand, using the spherically symmetric spacetime in Eq.\ref{le}, the Einstein's tensor $G_{\mu}^{\nu}$ is obtained as follows \begin{equation}
G_{\mu}^{\nu}=\operatorname{diag}\left[\frac{r f^{\prime}+f-1}{r^{2}}, \frac{r f^{\prime}+f-1}{r^{2}}, \frac{r f^{\prime \prime}+2 f^{\prime}}{2 r}, \frac{r f^{\prime \prime}+2 f^{\prime}}{2 r}\right].
\end{equation}

\begin{figure}[htp]
   \centering
    \includegraphics[scale=0.6]{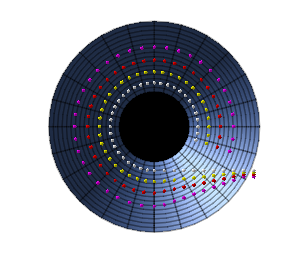}
        \includegraphics[scale=0.6]{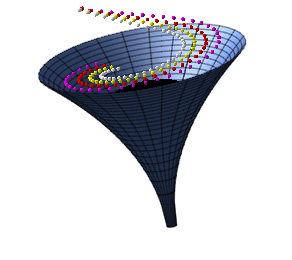}
    \caption{Illustration of the function $f(r)$ in the Cartesian coordinates. The first figure shows the top view and the second figure shows an angular view. The white circles represent the Schwarzschild case (SC); the yellow, red and purple circles correspond to different values of magnetic charge, $p = 1.4$, $p = 1.45$ and $p = 1.5$ respectively.}
    \label{fig:illus1}
\end{figure}

Then using the relation between the Einstein's tensor and energy momentum tensor
\begin{equation}
G_{\mu}^{\nu}=\kappa^{2} T_{\mu}^{\nu},
\end{equation}

where $
\kappa^{2}=8 \pi G
$, $t t$ components of the Einstein's field equations become
\begin{equation}
\frac{-1+r f^{\prime}(r)+f(r)}{r^{2}}=-\frac{1}{4 \pi \beta} \ln \left(\cosh ^{2}\left(\sqrt{\frac{\beta}{2}} \frac{p}{r^{2}}\right)\right)
\end{equation}
which gives the metric function $f(r)$
\begin{equation}f(r)=1-\frac{2 G M}{r}-\frac{2 G}{\beta r} \int^{r} x^{2} \ln \left[\cosh ^{2}\left(\sqrt{\frac{\beta}{2}} \frac{p}{x^{2}}\right)\right] \mathrm{~d} x, \end{equation}
that reduces in the weak field limit to the non-asymptotic behavior of magnetic Reissner-Nordstr\"om black hole \cite{Mazharimousavi:2021uki}
\begin{equation}
\label{fr}
f(r)=1-\frac{2 G M}{r}+\frac{G p^{2}}{r^{2}}-\beta \frac{G p^{4}}{60 r^{6}}+\beta^{2} \frac{G p^{6}}{810 r^{10}}-\mathcal{O}\left(\beta^{3}\right).
\end{equation}
This magnetic black hole solution will be the center of analyses henceforth. Fig. \ref{fig:illus1} illustrates the effect of the magnetic charge on incoming light rays. Note that the Reissner-Nordstr\"om metric belongs to the class of Stackel spaces. Geodesic equations can be exactly reintegrated (or reliably solved approximately) only in such spaces, because the Hamilton-Jacobi equation (for light rays - eikonal equation) admits in these spaces a complete separation of the variables and Obukhov present the method of complete separation of variables can be found in \cite{Obukhov}.

\begin{figure}[ht!]
   \centering
   \includegraphics[scale=0.4]{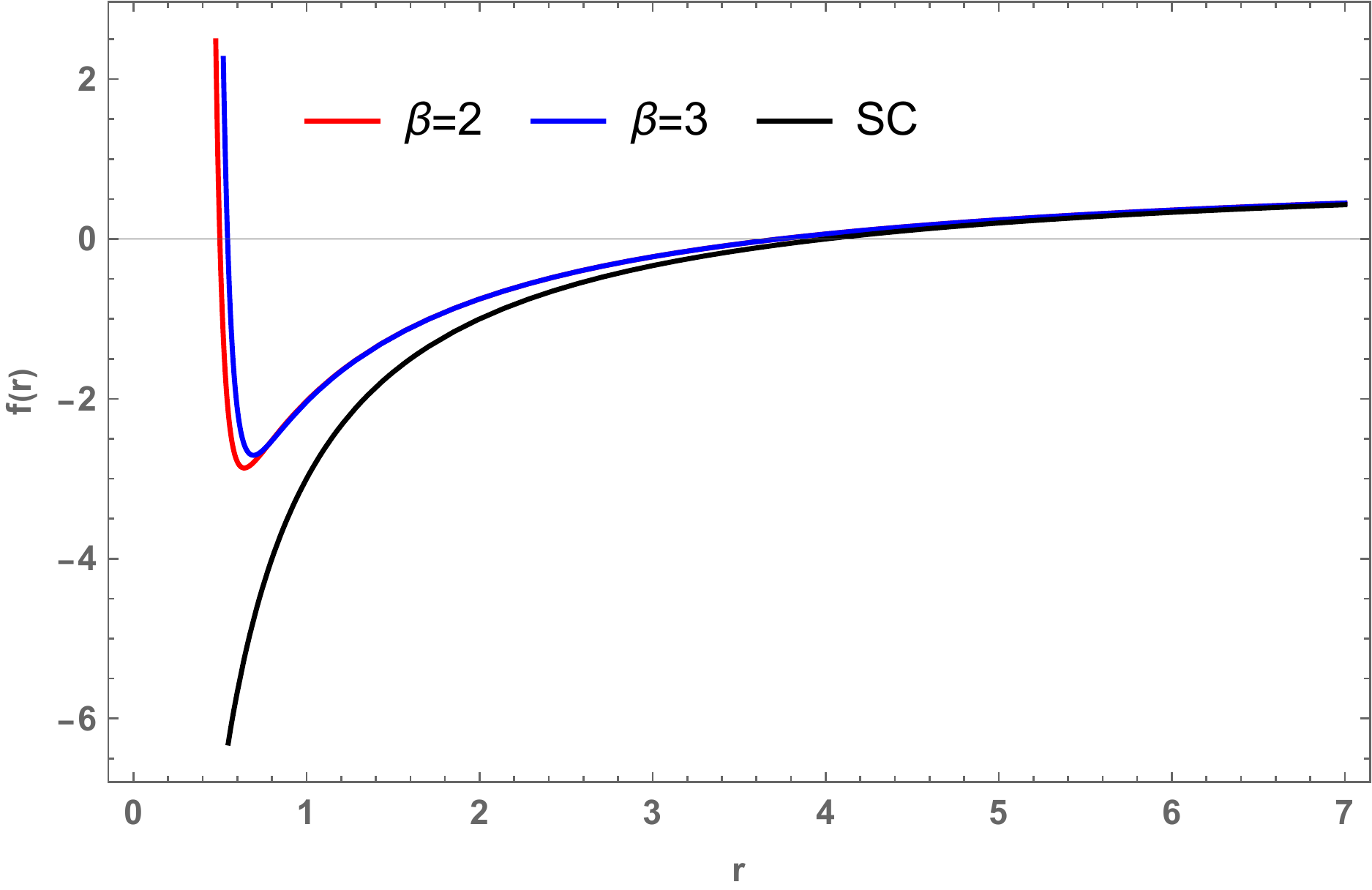}
    \caption{The lapse function $f(r)$ as a function of $r$ for $M=2$, $p=1$, $G=1$ and for the different values of $\beta$.}
    \label{fig:hr1}
\end{figure}

The event horizon radius $r_{+}$ of a black hole is the larger root of the above equation - where the outer horizon is located - with $f(r)=0$. As shown in Fig. \eqref{fig:hr1}, the number of horizons are dependant on the parameters of $f$.

\begin{figure}[ht!]
   \centering
    \includegraphics[scale=0.5]{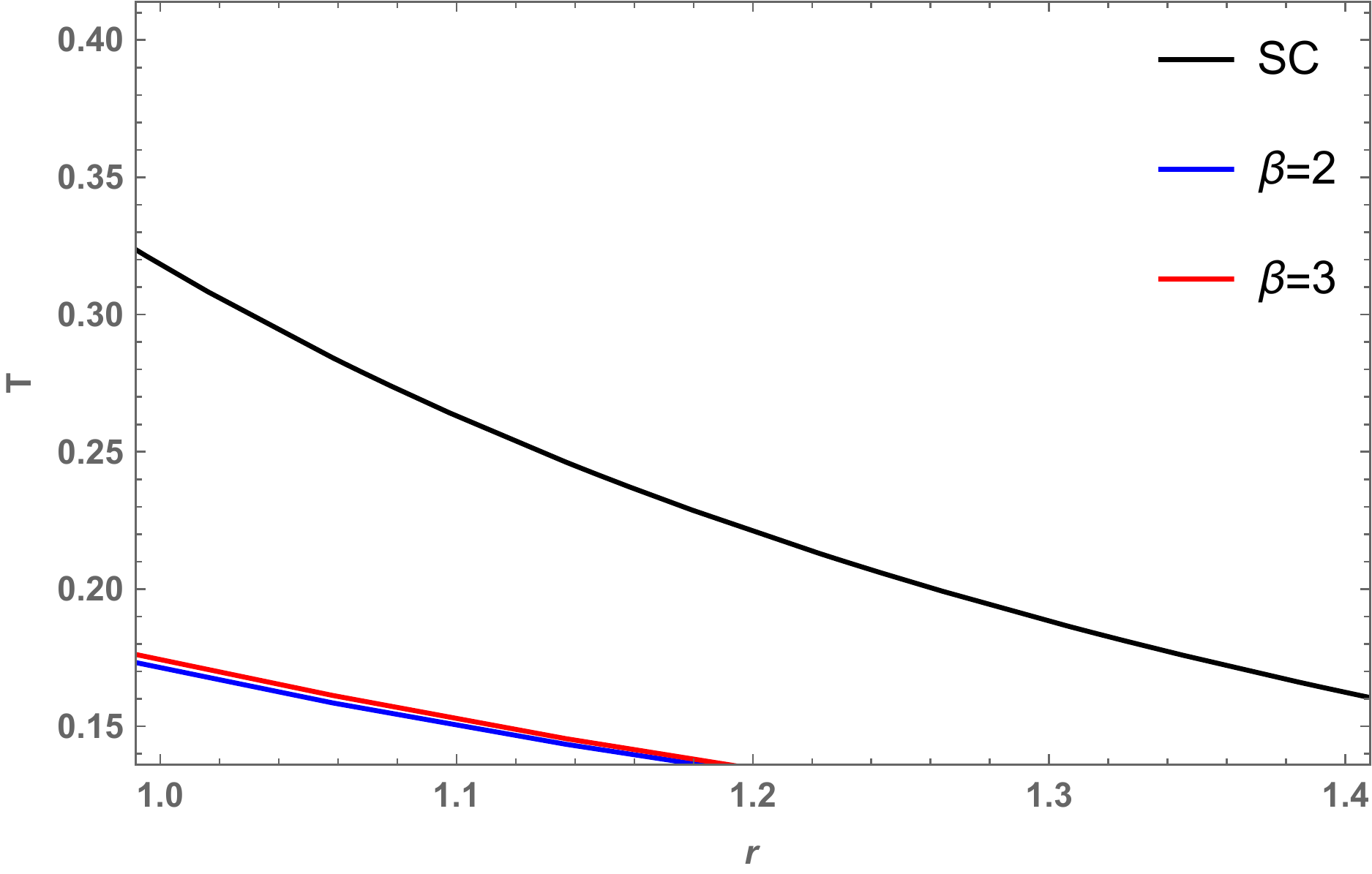}
        \includegraphics[scale=0.75]{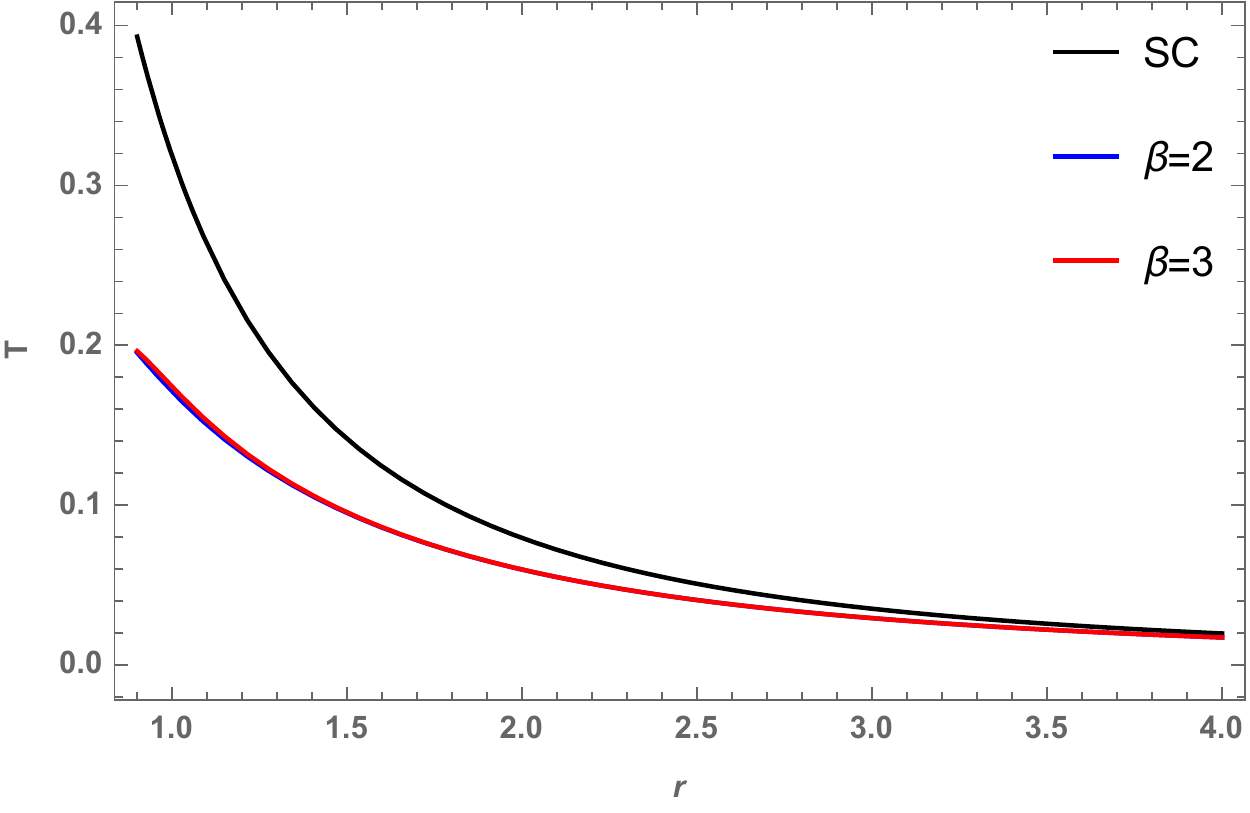}
    \caption{Hawking temperature $T$ (unit of 1/m in Geometrized unit) versus $r$ for $M=2$, $p=1$, and for the different values of $\beta$. 
The plot above is a zoom of the plot below.}
    \label{fig:temp}
\end{figure}

The 4-velocity is determined by
\begin{equation}
u=u^{t} \partial_{t},
\end{equation}
accompanied by the satisfying normalization condition
 \begin{equation}
1=u^{\mu} u_{\mu},
\end{equation}
where $u^{t}={1}/\sqrt{g_{t t}}$. Correspondingly, the particle acceleration $a_{p}^{\mu}$ is given by \cite{Wald:1984rg} 
 \begin{equation}a^{\mu}=-g^{\mu v} \partial_{v} \ln u^{t},\end{equation}
as the metric components exist as functions of $r$ and $\theta$. The surface gravity $(\kappa_{sg})$ is defined as
 \begin{equation}\kappa_{sg}=\lim _{r \rightarrow r_{h}} \frac{\sqrt{a^{\mu} a_{\mu}}}{u^{t}},\end{equation}
which helps in finding the black hole temperature
\begin{equation}
\kappa_{sg}=\left.\frac{1}{2} \partial_{r} f^{2}\right|_{r_{+}}, \quad T=\frac{\kappa_{sg}}{2 \pi},
\end{equation}

as shown in Fig.\ref{fig:temp} for different values of $\beta$.
The horizon area is plainly calculated as
\begin{equation}
A_{B H}=\int_{0}^{2 \pi} d \varphi \int_{0}^{\pi} \sqrt{-g} d \theta=4 \pi r_{+},
\end{equation}
giving the black hole entropy to be
\begin{equation}
S_{B H}=\frac{A_{B H}}{4}=\pi r_{+}.
\end{equation}

As for the thermodynamic properties of a black hole, the mass of a black hole described by $\left.g_{00}\right|_{r=r_{+}}=0$ is
\begin{equation}
M(r_{+})=\frac{r_{+}}{2 G}+\frac{\beta ^2 p^6}{1620 r_{+}^9}-\frac{\beta  p^4}{120 r_{+}^5}+\frac{p^2}{2 r_{+}}.
\end{equation}

At the limit of the Gauss-bonnet coupling and the magnetic parameters vanishing $\beta=p=0$, the black hole mass and temperature reduce to the Schwarzschild case, $M_{SC}={r_{+}}/{2G}$ and $T_{SC} =1/{8\pi M}$ respectively.

\section{Weak Deflection Angle using Gauss-Bonnet Theorem}
\label{wdagbt}
For the equatorial plane $\theta=\pi/2$, applying null geodesics to the line element in eq.~(\ref{le}) yields the optical metric
\begin{equation}
    \label{om}
    \mathrm{~d}t^2 = \frac{\mathrm{~d}r^2}{f(r)^2} + \frac{r^2}{f(r)} \mathrm{~d}\phi^2,
\end{equation}
where the determinant of the optical metric $g={r^2}/{f(r)^3}$.
In order to calculate the deflection angle while accounting for the optical geometry using a domain outside the light rays' trajectory, the Gauss-Bonnet theorem introduced by Gibbons and Werner is employed  \cite{Gibbons:2008rj}. The light rays are treated as spatial geodesics in the optical metric which induces a topological effect. The Gaussian curvature $\mathcal{K}$ is found to be proportional to the Ricci scalar computed from the non-zero Christoffel symbols as $\mathcal{K} = R/2$.

To calculate the deflection angle utilizing the optical Gaussian curvature, a non-singular region $\mathcal{D}_{R}$ with a boundary of $\partial \mathcal{D}_{R}=\gamma _{\tilde{g}}\cup C_{R}$ is selected. Alternatively, a non-singular domain outside of the light trajectory where the Euler characteristic $\chi (\mathcal{D}_{R})=1$ can also be chosen. For this region, the GBT can be stated as  \cite{Gibbons:2008rj} 
\begin{equation}
\label{gbtwb}
\iint\limits_{\mathcal{D}_{R}}\mathcal{K}\,\mathrm{~d}S+\oint\limits_{\partial \mathcal{%
D}_{R}}\kappa \,\mathrm{~d}t+\sum_{i}\theta _{i}=2\pi \chi (\mathcal{D}_{R}).
\end{equation}

As $R\rightarrow \infty $, the jump angles ($\theta _{\mathcal{O}}$, $\theta _{\mathcal{S}}$) can be equated to $\pi /2$ which says that the sum of the jump angles of the source $\mathcal{S}$, and of the observer $\mathcal{O}$ transpires as $\theta _{\mathit{O}} + \theta_{\mathit{S}} \rightarrow \pi $. Defining $\ddot{\gamma}$ as the unit acceleration vector, the geodesic curvature
\begin{equation}
\kappa =\tilde{g}\,\left(\nabla _{\dot{%
\gamma}}\dot{\gamma},\ddot{\gamma}\right),
\end{equation}
along with the unit speed condition $\tilde{g}(\dot{\gamma},\dot{\gamma}) = 1$ affects eq.~(\ref{gbtwb}) to become
\begin{equation}
\iint\limits_{\mathcal{D}_{R}}\mathcal{K}\,\mathrm{~d}S+\oint\limits_{C_{R}}\kappa \,%
\mathrm{~d}t\overset{{R\rightarrow \infty }}{=}\iint\limits_{\mathcal{D}%
_{\infty }}\mathcal{K}\,\mathrm{~d}S+\int\limits_{0}^{\pi +\hat{\alpha}}\mathrm{~d}\varphi
=\pi.
\end{equation}

Since $\gamma _{\tilde{g}}$ is a geodesic and $\kappa (\gamma _{\tilde{g}})=0$, choosing $C_{R}:=r(\varphi)=R=\text{const}$, the geodesic curvature becomes
\begin{equation}
\kappa (C_{R})=|\nabla _{\dot{C}_{R}}\dot{C}_{R}|,
\end{equation}
where, the radial part can be evaluated as
\begin{equation}
\left( \nabla _{\dot{C}_{R}}\dot{C}_{R}\right) ^{r}=\dot{C}_{R}^{\varphi
}\,\left( \partial _{\varphi }\dot{C}_{R}^{r}\right) +\tilde{\Gamma} _{\varphi
\varphi }^{r}\left( \dot{C}_{R}^{\varphi }\right) ^{2}. \label{12}
\end{equation}

While the first term vanishes, the second term is determined under the unit speed condition to be
\begin{equation}
\lim_{R\rightarrow \infty }\kappa (C_{R}) =\lim_{R\rightarrow \infty
}\left\vert \nabla _{\dot{C}_{R}}\dot{C}_{R}\right\vert \rightarrow \frac{1}{R}. 
\end{equation}
On the other hand, when the radial distance is very large
\begin{eqnarray}
\lim_{R\rightarrow \infty } \mathrm{~d}t&\to & R \, \mathrm{~d}\varphi,
\end{eqnarray}%

by combining the last two equations, $\kappa (C_{R})\mathrm{~d}t= \mathrm{~d}\,\varphi$. 

Then we use the straight-line approximation as $r = b/ \sin \phi$, where $b$ is the impact parameter. Thus the deflection angle can be determined by the Gibbons and Werner method through the Gauss-Bonnet theorem to be \cite{Gibbons:2008rj}

\begin{eqnarray}
\label{int0}
\hat{\alpha}=-\int\limits_{0}^{\pi}\int\limits_{{b}/{\sin \varphi}}^{\infty}\mathcal{K}\mathrm{~d}S,
\end{eqnarray}

where differential surface $\mathrm{~d}S = \sqrt{g}\mathrm{~d}r \mathrm{~d}\phi$.
Substituting eq.~(\ref{fr}) in eq.~(\ref{om}), the optical metric for a magnetic black hole using the new solution gives the Gaussian curvature evaluated from the non-zero Christoffel symbols to be

\begin{equation}
    \mathcal{K} \approx \frac{3 G^2 M^2}{r^4}+M \left(\frac{19 \beta  G^2 p^4}{30 r^9}-\frac{6 G^2 p^2}{r^5}-\frac{2 G}{r^3}\right)+\frac{p^4 \left(40 G^2 r^2-7 \beta  G\right)}{20 r^8}+\frac{3 G p^2}{r^4} + \mathcal{O}(p^6).
    \label{gauc}
\end{equation}

Ignoring the higher order terms, the above equations simplify to the asymptotic deflection angle given by
\begin{eqnarray}
    \hat{\alpha}\approx \frac{4 G M}{b}-\frac{3 \pi  G p^2}{4 b^2}+\frac{3 \pi  G^2 M^2}{4 b^2}+\frac{8 G^3 M^3}{3 b^3}-\frac{8 G^2 M p^2}{3 b^3}+\frac{7 \pi  \beta  G p^4}{384 b^6}\label{eq:GBTa}.
\end{eqnarray}

This reduces to the Schwarzschild case in the absence of the magnetic charge, $p=0$. 
Owing to the topological effects that exist as its salient features, the GBT method can be used in any asymptotically flat Riemannian optical metrics.

\begin{figure}[htp]
   \centering
    \includegraphics[scale=0.8]{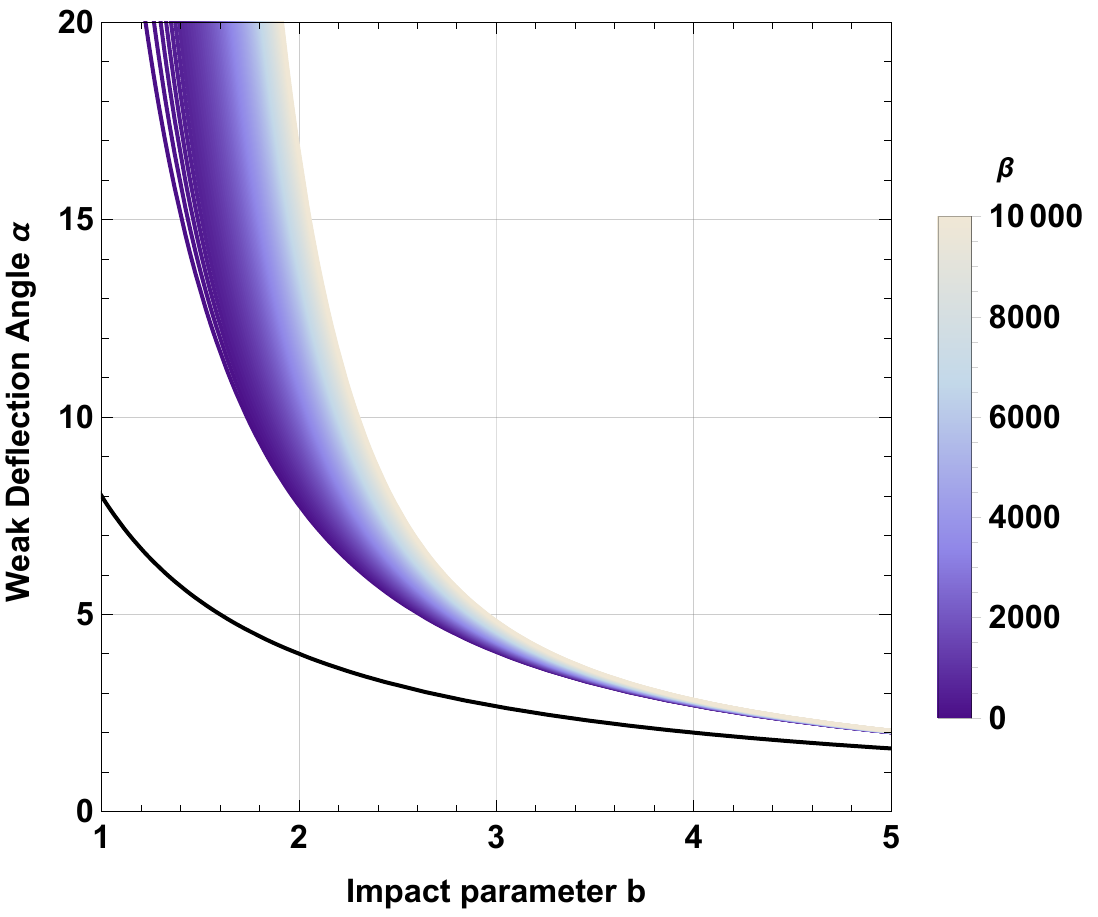}
        \includegraphics[scale=0.8]{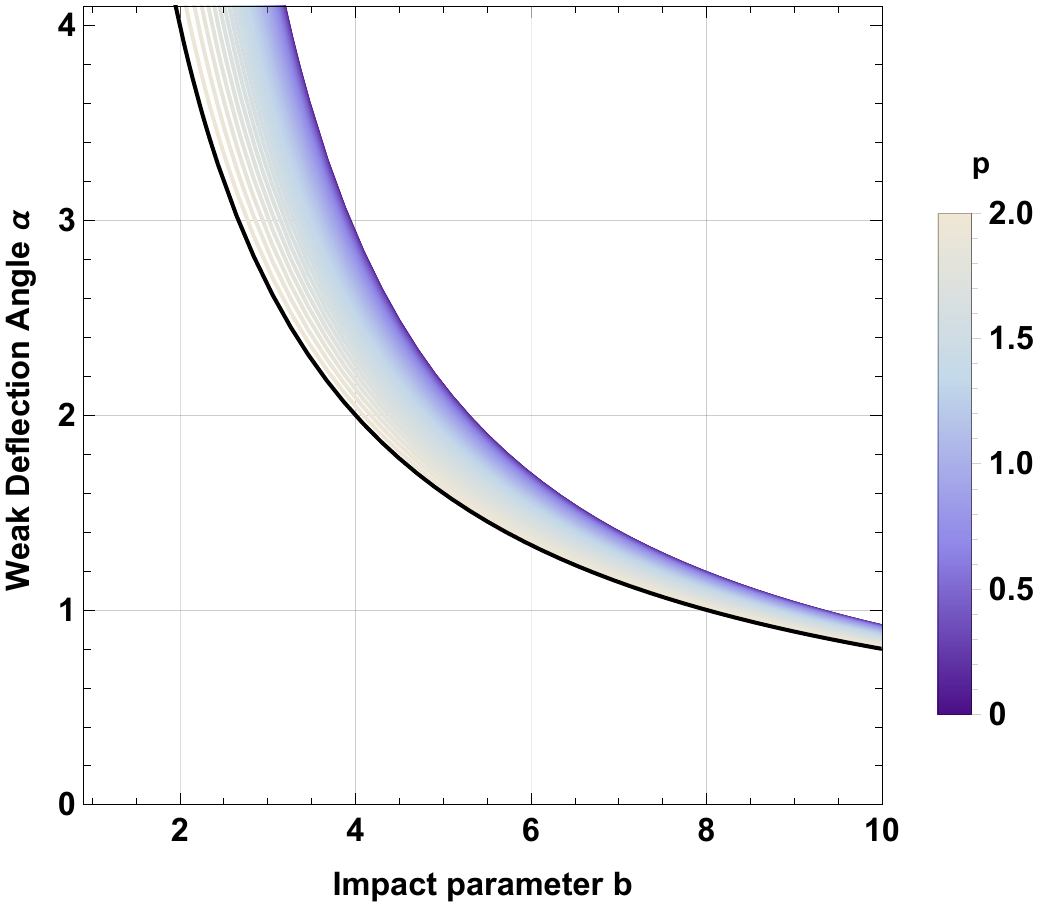}
    \caption{First figure shows the weak deflection angle $\hat{\alpha}$ versus impact parameter $b$ with $M=2$ and $p=1$ for different values of $\beta$. Second figure shows $\alpha$ versus $b$ with $M=2$ and $\beta=0.5$ for different values of $p$. The solid black line represents the Schwarzschild case.}
    \label{fig:lensing1}
\end{figure}

\begin{figure}[htp]
   \centering
    \includegraphics[scale=0.6]{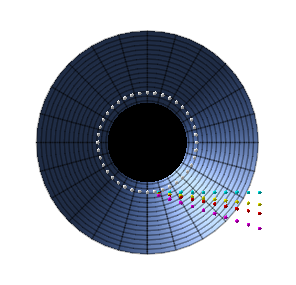}
        \includegraphics[scale=0.6]{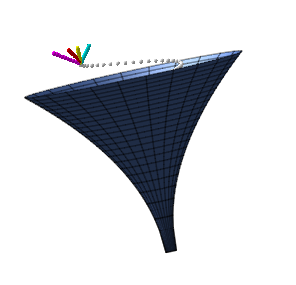}
    \caption{Illustration of the function $f(r)$ deflecting a ray of light for different values of $p$. The first figure shows the top view and the second figure shows an angular view. The while circles represent the ray of light; the cyan circles represent the Schwarzschild case; the yellow, red and purple circles correspond to different values of magnetic charge, $p = 1.4$, $p = 1.45$ and $p = 1.5$ respectively.}
    \label{fig:illus2}
\end{figure}

Fig. \ref{fig:lensing1} has been plotted to examine the effects of $\beta$ and $p$ more closely and separately. It is interesting to compare the variation scales of both figures: while $\beta$ does not show much difference for small changes, it tends to increase as $b$ increases, whereas, $p$ appears to be quite sensitive to any change and tends to decrease as $b$ increases. Therefore, for a given impact parameter, more the magnetic charge, more is the deflection. It is interesting to note that although both parameters tend to merge with the Schwarzschild case for high values of impact parameter, $\beta$ is close to the Schwarzschild case in the lower range, while $p$ is closer to the same in the higher range. This emphasizes on the nature of influence of $\beta$ and $p$ on the deflection angle and the inverse proportionality between each other. Fig. \ref{fig:illus2} illustrates the effect of the magnetic charge on the bending of an outgoing light ray.

 \subsection{\label{wdakpf} Calculation of weak deflection angle using Keeton-Petters formalism}

The Keeton-Petters formalism is a general PPN (post-post-Newtonian) approach \cite{Keeton:2005jd,Ruggiero:2016iaq} showing that \eqref{eq:GBTa} for the weak field deflection angle can be exactly obtained. The light propagation in the metric (\ref{le}) can be studied by following the general approach mentioned in \cite{Keeton:2005jd} for arbitrary spacetimes that are static and spherically symmetric. Specifically, the deflection of the light in the metric of the black hole and, more importantly, the corrections imposed to a few lensing observables owing to the non-linearity of the Lagrangian are focused.

Given the following metric
\beq
ds^{2}=-A(r)dt^{2}+B(r)dr^{2}+r^{2}d\Omega^{2} \ ,\label{eq:metricasimm}
\eeq
such that the coefficients are expressed in power series as
\begin{eqnarray}
A(r) & = & 1+2a_{1}\phi+2a_{2} \phi^{2}+2a_{3} \phi^{3}+... \label{eq:Apower} \ , \\
B(r) & = & 1-2b_{1}\phi+4b_{2} \phi^{2}-8b_{3} \phi^{3}+... \label{eq:Bpower}  \ , 
\end{eqnarray}
where $ \phi \equiv -M/r$ is defined as the Newtonian potential. Comparing this to the metric in eq.~(\ref{le}), $A(r)=f(r)$ and $B(r)=1/f(r)$. Setting $ P=p M$ leads to eq.~(\ref{le}), eq.~(\ref{eq:Apower}) and eq.~(\ref{eq:Bpower}) to yield
\beq
\left.\begin{array}{cc}a_{1}=G& a_{2}=G P^2/2 \ ,a_{3}=0, \\ b_{1}=G & b_{2}=\frac{1}{4} \left(4 G^2-G P^2\right). \end{array}\right. \label{eq:aibi}
\eeq

In General Relativity (GR), $\hat{\alpha}$ expanded as a series expresses the corrected deflection angle in the weak-field limit. Defining $r_{g} \equiv M$ as the gravitational radius of the source, the GR bending can be seen as a consequence of mass $M$ is $\hat \alpha_{GR}={4\, r_g}/{b}$. The derived bending angle from the aforementioned metric in eq.~(\ref{eq:metricasimm}) for first order of $M$ will be
\beq 
  \hat \varepsilon = A_1 \left(\frac{r_{g}}{b}\right) \ + \ A_2 \left(\frac{r_{g}}{b}\right)^2\ + \  O\left(\frac{r_{g}}{b}\right)^{3}. \label{eq:hatalpha}
\eeq
The coefficients $A_1,A_{2}$ do not depend on $M/b$. As for the coefficients $a_{1},b_{1},a_{2},\,\mathrm{and} \,b_{2}$, they become $A_{1}=2\left(a_{1}+b_{1} \right)$ and $A_{2}=\pi\left(2a_{1}^{2}-a_{2}+a_{1}b_{1}-\frac{b_{1}^{2}}{4}+b_{2} \right)$. With respect to eq.~(\ref{eq:aibi})
\beq
    A_{1}= 4 G\, ; A_{2}= \pi \left[\frac{1}{4} \left(4 G^2-G P^2\right)+\frac{11 G^2}{4}-\frac{G P^2}{2}\right].
    \label{eq:A1A2}
\eeq
Since the metric eq.~(\ref{eq:metricasimm}) is of the linear order of $M/r$, the contribution of the term proportional to $\displaystyle {M^{2}}/{b^{2}}$ can be neglected. Subsequently, the beding angle given by eq.~(\ref{eq:hatalpha}) evolves into
\begin{equation}
\begin{split}
    \hat{\varepsilon} =& \, \frac{4 G M}{b}-\frac{3 \pi  G p^2}{4 b^2} \quad\quad\quad, \mathrm{(or)} \\
    \hat{\varepsilon} =& \, \hat{\alpha}_{G R}\left(1-\frac{3 \pi  p^2}{16 b M}\right),
\end{split}
\end{equation}
which agrees with eq.~(\ref{eq:GBTa}) to the first order of $M/b$.

\begin{figure}
\centering
\includegraphics[scale=.55]{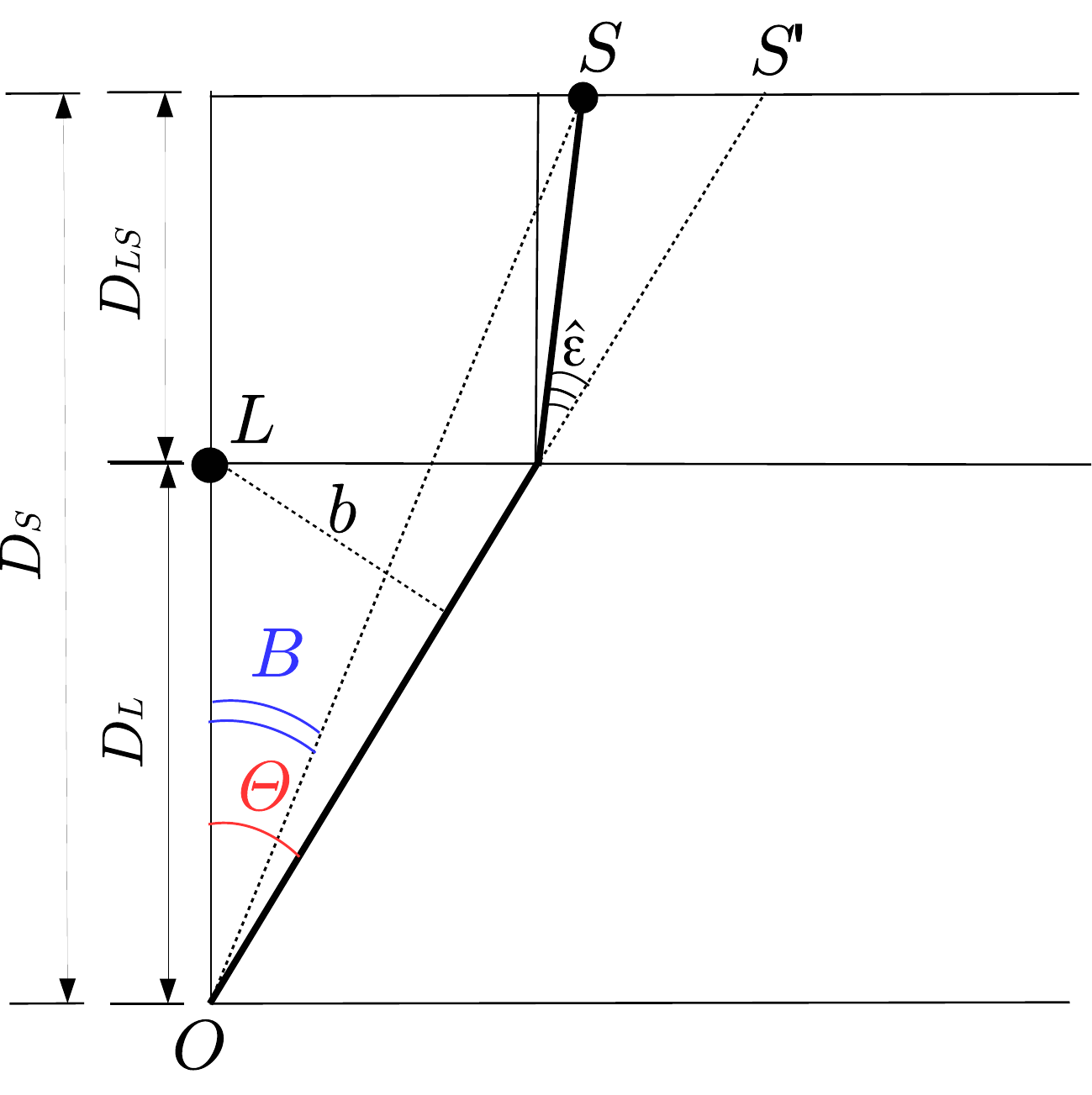}
\caption{Lens Geometry \cite{Ruggiero:2016iaq}.} \label{fig:lens}
\end{figure}

The lensing geometry is shown in Fig. \ref{fig:lens}. The angular position of the source is $B$ and that of the image is $\Theta$ with the bending angle $\hat{\varepsilon}$; $D$ is the distance and the subscripts $S$, $L$, and $LS$ correspond to observer-source, observer-lens, lens-source. An observer at point $O$ views the image at point $S$ of the source as though it was at $S'$. The location of the lens is at $L$. Applying the thin-lens approximation, all the light trajectories are approximated to straight lines. The impact parameter $b$ becomes a constant of motion in terms of the propagation of light: relative to an inertial observer at infinity, it is essentially the perpendicular distance from the point of the lens center at the asymptotic tangent to the line connecting the path of the ray of light to the observer. It is evident from Fig. \ref{fig:lens} that $b=D_{L} \sin \Theta$. Hence, the \textit{lens equation} can be derived by using elementary geometry to be
\beq
D_{S} \tan B = D_{S} \tan \Theta - D_{LS} \left[\,\tan \Theta-\tan \left(\Theta-\hat{\varepsilon} \right) \right].
\label{eq:lenseq}
\eeq
This can be used to determine $\Theta$ (angular position of the image) as a function of $B$ (angular position of the source) and $\hat{\varepsilon}$. Some underlying assumptions here are the lens being static and spherically symmetric, the source and the observer being asymptotically flat, and the light rays being propagated external to $r_{g}$ i.e. say, $r_0$ is the distance of nearest approach, then, $r_{0} \gg r_{g}$.

For small angles due to weak-lens approximation, the preceding lens equation  eq.~(\ref{eq:lenseq}) can be written as $ D_{S}\, B = D_{S} \,\Theta - D_{LS} \, \hat{\varepsilon}$. By employing the GR expression for both $\hat{\varepsilon}_{GR}={4\,M}/{b}$ and $ b = D_{L} \, \Theta $, the subsequent lens equation can be written as
\beq
D_{S}\, B = D_{S} \,\Theta - \frac{D_{LS}}{D_{L}}\frac{4 M}{\Theta}.
\label{eq:lensapprox}
\eeq

When $B=0$, the solution of this equation implies that the source, the lens and the observer are all aligned to be on the same line, giving rise to a characteristic angular scale called the \textit{Einstein angle}, $\theta_{E}$ expressed as
\beq
\theta_{E} = \sqrt{\frac{4M D_{LS}}{D_{L}D_{S}}}.
\label{eq:Eiangle}
\eeq
Sequentially, a characteristic length scale called the \textit{Einstein radius} $R_{E} = D_{L} \theta_{E}$ is also defined. All angular positions are scaled with respect to $\theta_{E}$ as
\beq
\zeta=\frac{B}{\theta_{E}}, \, \, \theta=\frac{\Theta}{\theta_{E}}, \, \, \textrm{(and setting) } \epsilon = \frac{\Theta_{M}}{\theta_{E}},
\eeq
where, $\Theta_{M}=\tan^{-1}(M/D_{L})$ is identified as the angle subtended by the lens' gravitational radius. The lensing observables are expanded in power series with the help of the parameter $\epsilon$. The lens equation from eq.~(\ref{eq:lenseq}), rendered similar to $\hat{\varepsilon}$ in the form of eq.~(\ref{eq:hatalpha}), for the position of the image becomes
\beq
\theta=\theta_{0}+\theta_{1}\epsilon+ \mathcal{O}(\epsilon^{2}), \label{eq:powerexpansionepsilon}
\eeq
where, $\theta_{0}$ is the position of the image in the weak-field limit solved as
\beq
0=-\beta+\theta_{0}-\frac{1}{\theta_{0}}. \label{eq:lens0}
\eeq

The images position, one of the observables, is accordingly obtained to be
\beq 
  \theta_0^\pm = \frac{1}{2}\left(\,\sqrt{ 4 + \beta^2} \pm |\beta| \right),
  \label{eq:imagepos0}
\eeq
where $\theta_{0}^{\pm}$ denotes the parity image with respect to the lens; the positive sign implies that the image lies on the same side as the source ($~\beta>0$) while the negative sign implies that the image lies on its other side opposite to the source ($~\beta<0$). Here, the second-order term becomes \cite{Keeton:2005jd}
\beq 
  \theta_1 = \frac{A_2}{A_1+4\theta_0^2}. \label{eq:theta1}
\eeq
Therefore, the image position up to the first order of $\epsilon$ is found to be
\beq
\theta=\theta_{0}+ \frac{A_2}{A_1+4\theta_0^2} \epsilon. \label{eq:thetatot1}
\eeq

Considering the values of $A_{1}$ and $A_{2}$ in eq.~(\ref{eq:A1A2}), the image position for the said black hole metric is written as
\beq
\theta = \theta_0+ \frac{3 \pi G \left(5 G M^2-p^2\right)}{16 M^2 \left(G+\theta_0^2\right)} \epsilon. \label{eq:thetatot1fT}
\eeq
The actual angular positions defined as  $\Theta=\theta \, \theta_{E}$ can be corrected as $\Theta_{1}=\theta_{1} \, \theta _{E} \, \epsilon$; for small angles, this is determined to be
\begin{equation}
\Theta_{1} \simeq \theta_{1} \frac{M}{D_{L}} \simeq \frac{3 \pi  G \left(5 G M^2-p^2\right)}{16 M^2} \frac{M}{D_{L}}. \end{equation}

The next observable is the magnification of an image $\mu$ at $\Theta$ that can be obtained with its sign in this formalism. Having the expression 

\beq \displaystyle \mu(\Theta) = \left[\frac{\sin B (\Theta)}{\sin \Theta} \, \frac{\mathrm{d}B (\Theta)}{\mathrm{d} \Theta} \right]^{-1}\eeq for general case, the series expansion in $\epsilon$ gives
\beq
\mu= \mu_{0}+\mu_{1} \epsilon + \mathcal{O}(\epsilon^{2}), \label{eq:mu1}
\eeq
where,
\beq
\mu_0 = \frac{16\theta_0^4}{16\theta_0^4-A_1^2} \quad \mathrm{and}\quad  \mu_1 = - \frac{16 A_2 \theta_0^3}{(A_1+4\theta_0^2)^3}. \label{eq:mu2}
\eeq

Once again, taking the values of $A_{1}$ and $A_{2}$ from eq.~(\ref{eq:A1A2}) into account for the metric in question

\beq
\mu_0 = \frac{\theta_0^4}{\theta_0^4-G^2}, \quad \quad  \mu_1 = -\frac{3 \pi  G \theta_0^3 \left(5 G-P^2\right)}{16 \left(G+\theta_0^2\right)^3}. \label{eq:mu3}
\eeq

Naturally, $\mu>0$ corresponds to $\theta^{+}$, the positive parity image, and $\mu<0$ to $\theta^{-}$, the negative parity image. The sign of $P$ is seen to influence the magnification; if $P^2>5G$, then $\mu_{1}<0$ and the positive-parity image appears faint while the negative-parity image is bright. Technically, this condition can be utilized as a test in observation for the theoretical quantity $P$. Since the total magnification is not altered to the first-order of $\epsilon$ when the second-order term is rather proportional to $A_{2}^{2}$, it is found to be null in this approximation.

The last observable is the time delay, which is defined as the travelling time difference (due to the lens) between the actual time that light takes and the time that it would take if there were no lens, and it is given by
\beq
  \frac{\tau}{\tau_E} =
    \frac{1}{2}
    \left[ a_1 + \beta^2 - \theta_0^2 - \frac{a_1+b_1}{2} \ln \left(
    \frac{D_L\,\theta_0^2\,\theta_E^2}{ 4\,D_{LS}} \right) \right]
\ + \ \frac{\pi}{16\,\theta_0}
    \Bigl( 8 a_1^2 - 4 a_2 + 4 a_1 b_1 - b_1^2 + 4 b_2 \Bigr)\,\epsilon
\ + \ \mathcal{O}({\epsilon^{2}}),   \label{eq:delay111}
\eeq
which becomes
\beq
   \frac{\tau}{\tau_E} =
    \frac{1}{2}
    \left[ 4G + \beta^2 - \theta_0^2 - \frac{5G}{2} \ln \left(
    \frac{D_L\,\theta_0^2\,\theta_E^2}{ 4\,D_{LS}} \right) \right]
\ + \ \frac{\pi}{16\,\theta_0}
    \Bigl( 15 G^2 - 3 \frac{G p^2}{2 M^2}  \Bigr)\,\epsilon
\ + \ \mathcal{O}({\epsilon^{2}}). \label{eq:timedelay1}
\eeq
for this case under scrutiny. In physical units, $\tau_{E}=4{GM}/{c^{3}}$ while in natural units, the gravitational constant $G$ and the speed of light $c$ are taken to be unity and thus, $\tau_{E}=4M$ is obtained for the regarded system in the natural time-scale. The differential time delay can be calculated between the negative and the positive parity images as
\beq \label{eq-PPN-dtau}
  \Delta \tau = \Delta\tau_0 \ + \ \vep\,\Delta\tau_1 \ + \mathcal{O}(\epsilon^{2}),
\eeq
where
\begin{eqnarray}
  \Delta\tau_0 &=& \tau_E
    \left[ \frac{(\theta_0^{-})^{-2} - (\theta_0^{+})^{-2}}{2}
    - \frac{a_1+b_1}{2}\,\ln\left(\frac{\theta_0^{-}}{\theta_0^{+}}\right)
    \right] , \\
  \Delta\tau_1 &=& \tau_E\ \frac{\pi}{16}
    \Bigl( 8 a_1^2 - 4 a_2 + 4 a_1 b_1 - b_1^2 + 4 b_2 \Bigr)
    \frac{(\theta_0^{+}-\theta_0^{-})}{\theta_0^{+} \theta_0^{-}}.
\end{eqnarray}

Applying the values from eq.~(\ref{eq:aibi}) for $a_{1},b_{1},a_{2},$ and $b_{2}$ for the RR metric, the above equations become
\begin{eqnarray}
   \Delta\tau_0 &=& \tau_E
    \left[ \frac{(\theta_0^{-})^{-2} - (\theta_0^{+})^{-2}}{2}
    - G\,\ln\left(\frac{\theta_0^{-}}{\theta_0^{+}}\right)
    \right], \\
  \Delta\tau_1 &=& \tau_E\ \frac{\pi}{16}
    \left( 15 G^2 - \frac{3G p^2}{2 M^2} \right)
    \left(\frac{\theta_0^{+}-\theta_0^{-}}{\theta_0^{+} \theta_0^{-}}\right).
    \label{eq:tau1}
\end{eqnarray}

The time delay corrected to first-order gives an order of magnitude

\begin{equation}
    \Delta \tau_{1} \epsilon \simeq \frac{\tau_{E}}{16} \Bigl( 15 G^2 - \frac{3G p^2}{2 M^2}  \Bigr) \pi \epsilon.
\end{equation}

\section{Weak deflection angle of massive particles}
\label{wdajm}
For a static and spherically symmetric (SSS) spacetime, the general form
\begin{equation}
    \mathrm{~d}s^2=g_{\mu \nu}\mathrm{~d}x^{\mu}\mathrm{~d}x^{\nu} =-A(r)\mathrm{~d}t^2+B(r)\mathrm{~d}r^2+C(r)\mathrm{~d}\Omega^2,
\end{equation}
can be re-written for the Jacobi metric as
\begin{align}
    \mathrm{~d}l^2&=g_{ij}\mathrm{~d}x^{i}\mathrm{~d}x^{j}=\left(E^2-m^2 A(r)\right) \left(\frac{B(r)}{A(r)}\mathrm{~d}r^2+\frac{C(r)}{A(r)}\mathrm{~d}\Omega^2\right),
\end{align}
where, $E$ is the particle energy per unit mass $m$ and  $\mathrm{~d}\Omega^2=\mathrm{~d}\theta^2+r^2\sin^2\phi$ is the line element of the unit two-sphere. The Jacobi metric can be utilized to derive the radius of the circular photon orbit using a geometric method for a particle in the equatorial plane. The Jacobi metric $\mathrm{~d}l^2$ obviously reduces to the optical metric $\mathrm{~d}t^2$ for null particles with $E=1$ and $m=0$ (A non-zero mass term implies that the particle under consideration is a massive particle and thus, not travelling at the speed of light $c$). An asymptotic observer perceives one of the constants of motion, the particle energy, far away from the black hole as
\begin{equation}
    E = \frac{m}{\sqrt{1-v^2}},
\end{equation}
where, $v$ is the particle velocity prevailing as a fraction of $c$. In the equatorial plane $\theta = \pi/2$, the Jacobi metric is \cite{Li:2019vhp}
\begin{equation}
    \mathrm{~d}l^{2}=m^{2}\left(\frac{1}{1-v^{2}}-A(r)\right) \left(\frac{B(r)}{A(r)} \mathrm{~d}r^{2}+\frac{C(r)}{A(r)} \mathrm{~d}\phi^{2}\right),
\end{equation}
maintaining the generality. The determinant of the above metric is determined to be
\begin{equation}
    g=m^4 B(r)C(r) ~\left[\frac{A(r)\,(v^2-1)+1}{A(r)\,(v^2-1)}\right]^2.
\end{equation}

Using GBT to find the weak deflection angle,
\begin{eqnarray}\label{int01}
\hat{\alpha}=-\int\limits_{0}^{\pi}\int\limits_{{b}/{\sin \varphi}}^{\infty}\mathcal{K}\mathrm{~d}S,
\end{eqnarray}
for the Jacobi metric defined above, the weak deflection angle of massive particles is written as
\begin{eqnarray}
    \hat{\alpha}=\frac{\pi  \beta  G p^4}{384 b^6}+\frac{\pi  \beta  G p^4}{64 b^6 v^2}+\frac{16 G^3 M^3}{3 b^3 v^6}-\frac{40 G^3 M^3}{3 b^3 v^4}+\frac{10 G^3 M^3}{b^3 v^2}+\frac{2 G^3 M^3}{3 b^3}+\frac{4 G^2 M p^2}{b^3 v^4}-\frac{6 G^2 M p^2}{b^3 v^2} \notag\\-\frac{2 G^2 M p^2}{3 b^3}-\frac{\pi  G^2 M^2}{b^2 v^4}+\frac{3 \pi  G^2 M^2}{2 b^2 v^2}+\frac{\pi  G^2 M^2}{4 b^2}-\frac{\pi  G p^2}{2 b^2 v^2}-\frac{\pi  G p^2}{4 b^2}+\frac{2 G M}{b v^2}+\frac{2 G M}{b}.
\end{eqnarray}

\begin{figure}[htp]
   \centering
    \includegraphics[scale=0.8]{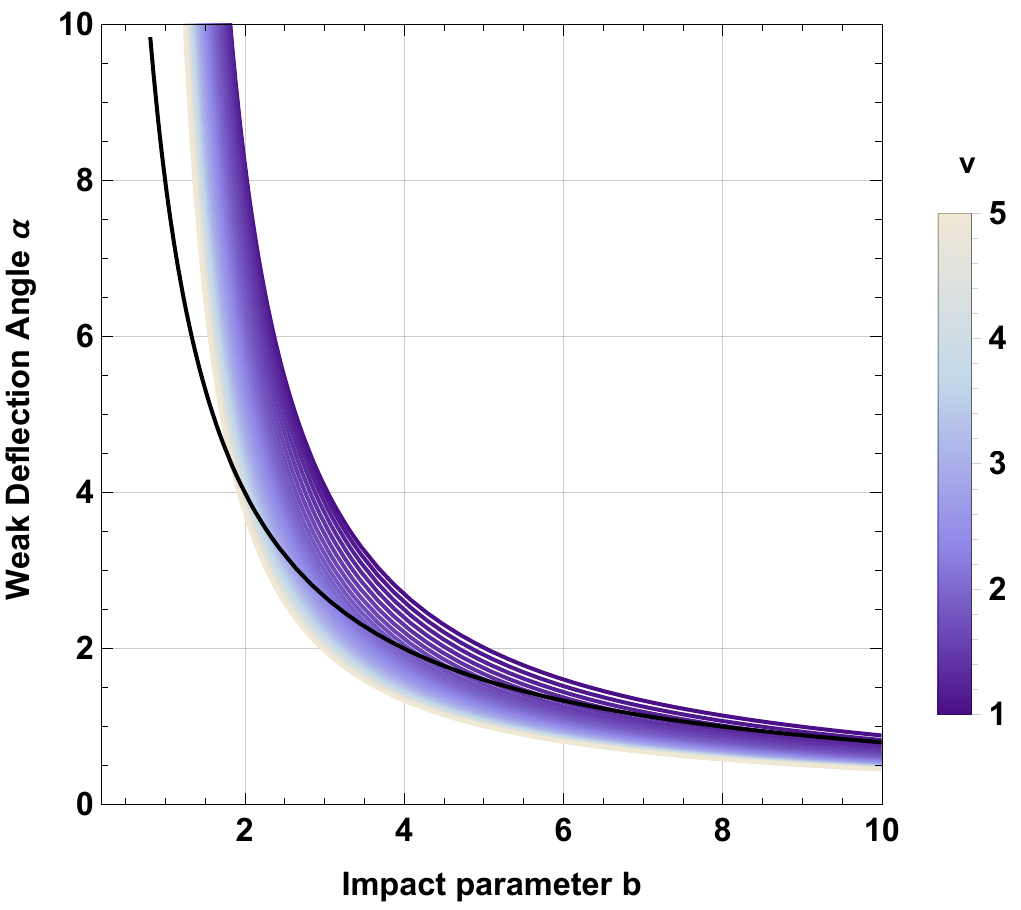}
        \caption{Figure shows the weak deflection angle $\hat{\alpha}$ versus impact parameter $b$ for $M=2$, $p=1$ and $\beta=500$ for different values of $v$. The solid black line corresponds to a photon with $v=c=1$.}
    \label{fig:lensing3}
\end{figure}

For null particles, $v=1$ which in turn gives
\begin{equation} \label{ewda2}
    \hat{\alpha}= \frac{4 G M}{b}-\frac{3 \pi  G p^2}{4 b^2}+\frac{7 \pi  \beta  G p^4}{384 b^6}+\frac{3 \pi  G^2 M^2}{4 b^2}-\frac{8 G^2 M p^2}{3 b^3}+\frac{8 G^3 M^3}{3 b^3},
\end{equation}
as expected. Furthermore, when there are no charge, the Schwarzschild case $\hat{\alpha}=4M/b$ is recovered. In Fig. \ref{fig:lensing3}, the case of the deflection angle due to a massive particle is plotted against the impact parameter. The parameter $v$ is observed to reduce the value of $\hat{\alpha}$, approaching the Schwarzschild case as $b/M \to \infty$.

\section{Weak Deflection Angle in plasma medium and dark matter}
\label{wdapdm}

\subsection{Weak Deflection Angle in plasma medium}
Another non-trivial factor influencing a gravitational lens is plasma. Refraction arises in a plasma medium causing more deflection. It is particularly eminent in the radio regime and is described by the refractive index which pursues to acquire auxiliary components.

To encompass the effects of plasma, let $v$ be the velocity of light travelling through hot, ionized gas. The refractive index, $n(r) \equiv c/v$ 
for the magnetic black hole in question is expressed by \cite{Crisnejo:2018uyn}
\begin{equation}
    n(r)=\sqrt{1-\frac{\omega_e^2}{\omega_\infty^2} \left[1-\frac{2 G M}{r}+\frac{G p^{2}}{r^{2}}- \frac{G \beta\,p^{4}}{60 r^{6}}+ \frac{G \beta^2 p^{6}}{810 r^{10}}\right]},
\end{equation}
where $c=1$, $\omega_{e}$ is the electron plasma frequency and $\omega_{\infty}$ is the photon frequency measured by an observer at infinity. The line element in equation (\ref{le}) is re-written as
\begin{equation}
    \mathrm{~d} \sigma ^ { 2 } = g _ { i j } ^ { \mathrm { opt } } \mathrm{~d} x ^ { i } \mathrm{~d} x ^ { j } = \frac { n ^ { 2 } ( r ) } { f(r) } \left[ \frac{\mathrm{~d} r ^ { 2 }}{f(r)} + r^2 \mathrm{~d} \phi ^ { 2 }\right].
\end{equation}

This yields the optical Gaussian curvature to be
\begin{equation}
    \begin{split}
        \mathcal{K} \approx & -\frac{12 G^3 M^3 \omega _e^2}{r^5 \omega _{\infty }^2}-\frac{14 \beta  G^3 M^2 p^4 \omega _e^2}{5 r^{10} \omega _{\infty }^2}+\frac{32 G^3 M^2 p^2 \omega _e^2}{r^6 \omega _{\infty }^2}-\frac{23 G^3 M p^4 \omega _e^2}{r^7 \omega _{\infty }^2}+\frac{12 G^2 M^2 \omega _e^2}{r^4 \omega _{\infty }^2} \\
        & +\frac{27 \beta  G^2 M p^4 \omega _e^2}{10 r^9 \omega _{\infty }^2}-\frac{26 G^2 M p^2 \omega _e^2}{r^5 \omega _{\infty }^2} +\frac{10 G^2 p^4 \omega _e^2}{r^6 \omega _{\infty }^2}-\frac{3 G M \omega _e^2}{r^3 \omega _{\infty }^2}-\frac{13 \beta  G p^4 \omega _e^2}{20 r^8 \omega _{\infty }^2}+\frac{5 G p^2 \omega _e^2}{r^4 \omega _{\infty }^2} \\
        & +\frac{3 G^2 M^2}{r^4}+\frac{19 \beta  G^2 M p^4}{30 r^9}-\frac{6 G^2 M p^2}{r^5}+\frac{2 G^2 p^4}{r^6}-\frac{2 G M}{r^3}-\frac{7 \beta  G p^4}{20 r^8}+\frac{3 G p^2}{r^4}.
    \end{split}
\end{equation}

With the GBT becoming

\begin{equation}
\lim_{\rho\rightarrow\infty}\int_{0}^{\pi+\hat{\alpha}}\left[\kappa\frac{\mathrm{~d}\sigma}{\mathrm{~d}\varphi}\right]\bigg|_{C_{\rho}}\mathrm{~d}\varphi =\pi-\lim_{\rho \rightarrow \infty }\int \int_{D_{\rho}} \mathcal{K} \mathrm{~d}S,
\label{alpha-bonnet}
\end{equation}

where,
\begin{equation}
\ \kappa (C_\rho) \mathrm{~d} t = \mathrm{~d} \varphi,
\end{equation}
for very large radial distances. Therefore, for this profile of number density and the physical metric that indicates an asymptotically Euclidean optical metric, it can be seen that
\begin{equation}
\lim_{\rho\rightarrow\infty}\kappa\frac{\mathrm{~d}\sigma}{\mathrm{~d}\varphi}\bigg|_{C_{\rho}}=1,
\label{khomons}
\end{equation}
as expected. For the linear order of $M$, using eq.~(\ref{alpha-bonnet}) in the limit of $\rho\rightarrow\infty$, and taking $\gamma$, the geodesic curve that is approximated by its flat Euclidean counterpart parameterized as $\rho=b/\sin\varphi$, in the physical spacetime to give

\begin{equation}
\hat{\alpha}=-\lim_{\rho\rightarrow\infty}\int_{0}^{\pi}\int_{{b}/{\sin\varphi}}^{\rho}\mathcal{K} \mathrm{~d}S.
\label{alpha1}
\end{equation}

The deflection angle of the black hole in a medium for the leading order terms is calculated non-trivially to be
\begin{eqnarray}
\hat{\alpha}=\frac{\pi  \beta  G p^4 \omega _e^2}{64 b^6 \omega _{\infty }^2}+\frac{7 \pi  \beta  G p^4}{384 b^6}-\frac{2 G^3 M^3 \omega _e^2}{3 b^3 \omega _{\infty }^2}+\frac{2 G^2 M p^2 \omega _e^2}{b^3 \omega _{\infty }^2}+\frac{8 G^3 M^3}{3 b^3}-\frac{8 G^2 M p^2}{3 b^3} \notag \\-\frac{\pi  G^2 M^2 \omega _e^2}{2 b^2 \omega _{\infty }^2}-\frac{\pi  G p^2 \omega _e^2}{2 b^2 \omega _{\infty }^2}+\frac{3 \pi  G^2 M^2}{4 b^2}-\frac{3 \pi  G p^2}{4 b^2}+\frac{2 G M \omega _e^2}{b \omega _{\infty }^2}+\frac{4 G M}{b},\label{phs}
\end{eqnarray}
which agrees with the conventional results in the limit where it's presence is insignificant ($n_0=1$), reducing to the familiar expression in vacuum, $\hat{\alpha}=4{M}/{b}$. Note that GBT is exhibiting a topological effect partially.

\begin{figure}[htp]
   \centering
    \includegraphics[scale=0.8]{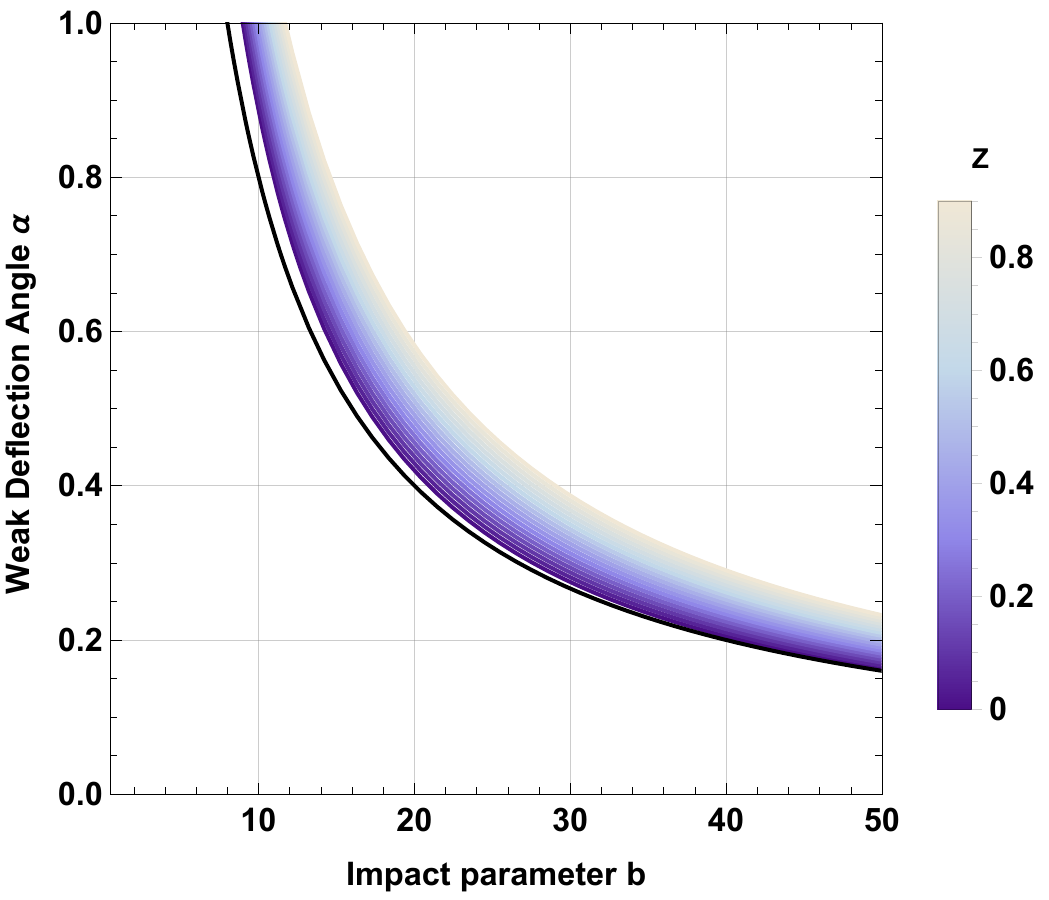}
        \caption{Figure shows the weak deflection angle $\hat{\alpha}$ versus impact parameter $b$ for $M=2$, $p=1$ and $\beta=500$ for different values of refraction parameter, $Z \equiv {\omega_e^2}/{\omega_\infty^2}$. The solid black line represents the Schwarzschild case.} 
    \label{fig:lensing4}
\end{figure}

Fig. \ref{fig:lensing4} is plotted to study the effects of plasma. For obvious reasons, the range of the refraction parameter, $Z \equiv {\omega_e^2}/{\omega_\infty^2}$ is set to be $ \leq Z \leq 0.9$. Evidently, plasma appears to increase the deflection angle for a given value of $b$: this is expected as the bending that occurs in this case is due to both gravity and refraction, resulting in a distinctly additional component.

\subsection{Weak Deflection Angle in dark matter medium}
The refractive index is defined for the dark matter medium as \cite{Latimer:2013rja,Ovgun:2018oxk,Ovgun:2020yuv}

\begin{equation}
n(w) = 1 + B u + v w^2,
\label{dm}
\end{equation}
where, $w$ is the frequency of light, $B \equiv { \rho_0 }/{ 4 m ^ { 2 }w ^ { 2 } } $ with $\rho_0$ as the mass density of the scattered particles of dark matter, $u= - 2 \varepsilon ^ { 2 } e ^ { 2 }$ with $\varepsilon$ as the charge of the scatterer in units of $e$, and $v \geq 0$. The order of terms in $\mathcal{O} \left(w^2\right)$ and higher are associated with the polarizability of the dark matter particle. This is the anticipated refractive index for an optically inactive medium. The order of $w^{-2}$ corresponds to a dark matter candidate that is charged and $w^{2}$ to a dark matter candidate that is neutral. Additionally, a linear term in $w$ is a possibility when parity and charge-parity asymmetries exist. Consequently, the Gaussian curvature is obtained as
\begin{eqnarray}
\mathcal{K} \approx \frac{3 G^2 M^2}{r^4 \left(B u+v w^2+1\right)^2}+\frac{19 \beta  G^2 M p^4}{30 r^9 \left(B u+v w^2+1\right)^2}-\frac{6 G^2 M p^2}{r^5 \left(B u+v w^2+1\right)^2}+\frac{2 G^2 p^4}{r^6 \left(B u+v w^2+1\right)^2} \notag \\-\frac{2 G M}{r^3 \left(B u+v w^2+1\right)^2}-\frac{7 \beta  G p^4}{20 r^8 \left(B u+v w^2+1\right)^2}+\frac{3 G p^2}{r^4 \left(B u+v w^2+1\right)^2}+\mathcal{O}(M^3),
\end{eqnarray}
prompting the deflection angle to be
\begin{eqnarray}
\hat{\alpha}= -\frac{7 \pi  \beta  B G p^4 u}{192 b^6}+\frac{7 \pi  \beta  B G p^4 u v w^2}{64 b^6}+\frac{7 \pi  \beta  G p^4}{384 b^6}-\frac{7 \pi  \beta  G p^4 v w^2}{192 b^6}+\frac{16 B G^2 M p^2 u v w^2}{b^3}\notag \\-\frac{16 B G^2 M p^2 u}{3 b^3} -\frac{16 G^2 M p^2 v w^2}{3 b^3}+\frac{8 G^2 M p^2}{3 b^3}-\frac{9 \pi  B G^2 M^2 u v w^2}{2 b^2}+\frac{3 \pi  B G^2 M^2 u}{2 b^2}\notag \\-\frac{9 \pi  B G p^2 u v w^2}{2 b^2}+\frac{3 \pi  B G p^2 u}{2 b^2} +\frac{3 \pi  G^2 M^2 v w^2}{2 b^2}-\frac{3 \pi  G^2 M^2}{4 b^2}+\frac{3 \pi  G p^2 v w^2}{2 b^2}\notag \\-\frac{3 \pi  G p^2}{4 b^2}+\frac{24 B G M u v w^2}{b}-\frac{8 B G M u}{b}-\frac{8 G M v w^2}{b}+\frac{4 G M}{b}.
\label{da1}
\end{eqnarray}

Therefore, the dark matter medium can be seen to affect the deflection angle by inducing small deflections in comparison to the standard case of a Schwarzschild black hole.

\begin{figure}[htp]
   \centering
    \includegraphics[scale=0.8]{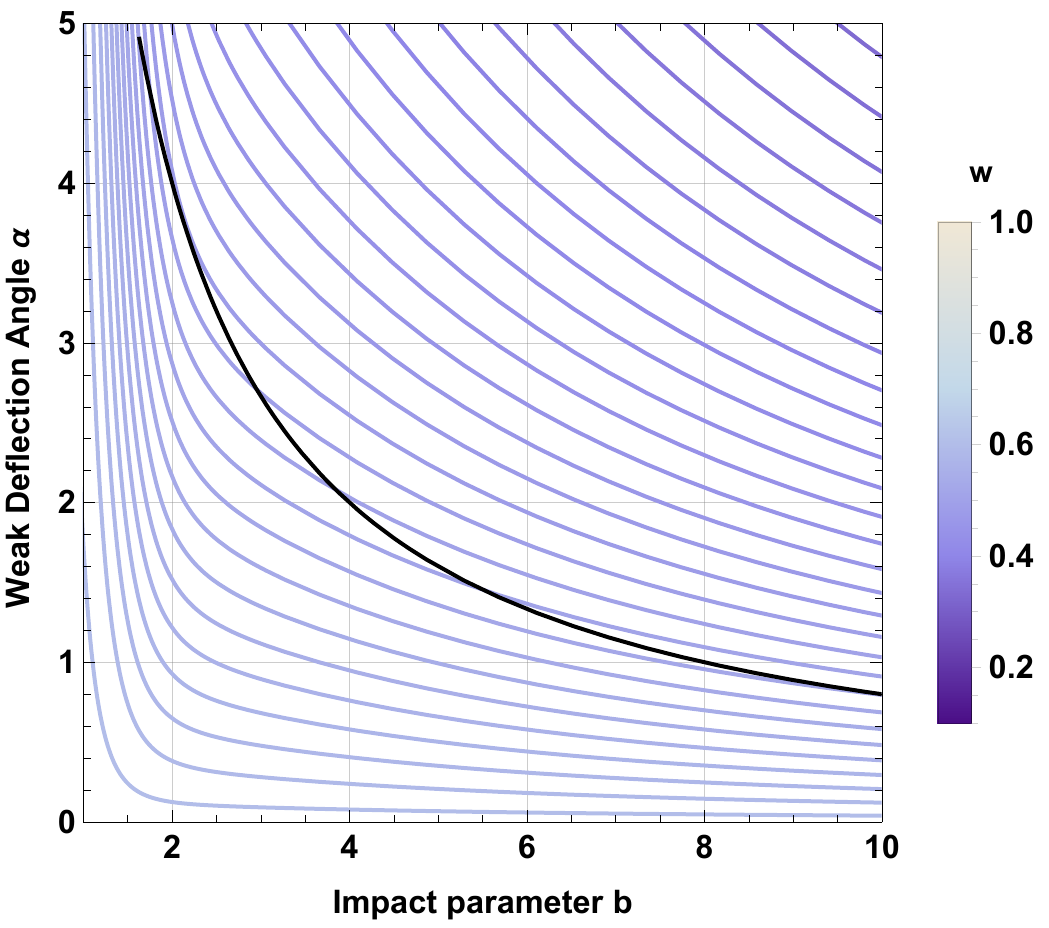}
        \caption{Figure shows the weak deflection angle $\hat{\alpha}$ versus impact parameter $b$ for $M=2$, $p=1$, $\beta=500$, $u=-2$, $\rho_0=1$ and $v=m=1$ for different values of $w$. The solid black line represents the case of vacuum containing no dark matter.}
    \label{fig:lensing5}
\end{figure}

Fig. \ref{fig:lensing5} shows the variation of the bending angle with respect to the impact parameter in the presence of dark matter. Its effect is more spaced out than the other cases that were examined -- dark matter is observed to impact the deflection angle drastically. Apparently, low values of the dark matter parameter $w$ is seen to produce a high deflection, whereas, high values of $w$ gives low deflection.

\section{Shadow of the black hole}
\label{mcsbh}
Here, the shadow of a black hole is analysed and the effect of the magnetic charge on the shadow cast is studied. A black hole shadow represents the interior of the so-called apparent boundary or the critical curve, the latter being such that the light rays that are a part of it approaches a bound orbit of photons asymptotically when a distant observer traces it back to the black hole. The Hamilton-Jacobi approach in the equatorial plane $\theta={\pi}/{2}$ for a photon is expressed as \cite{Perlick:2021aok}
\begin{equation}
H=\frac{1}{2} g^{\mu v} p_{\mu} p_{v}=\frac{1}{2}\left(\frac{L^{2}}{r^{2}}-\frac{E^{2}}{f(r)}+\frac{\dot{r}^{2}}{f(r)}\right)=0 \label{hj},
\end{equation}
where, $p_{\mu}$ is the momentum of the photon, $L = p_\phi$ is its angular momentum, $E = -p_t$ is its energy, and $\dot{r}\equiv \partial H / \partial p_{r}$ . Using the above equation, a complete dynamics with an effective potential $V$ is described by
\begin{equation}
V+\dot{r}^{2}=0, \quad V=f(r)\left(\frac{L^{2}}{r^{2}}-\frac{E^{2}}{f(r)}\right).
\end{equation}

\begin{figure}[htp]
   \centering
    \includegraphics[scale=0.8]{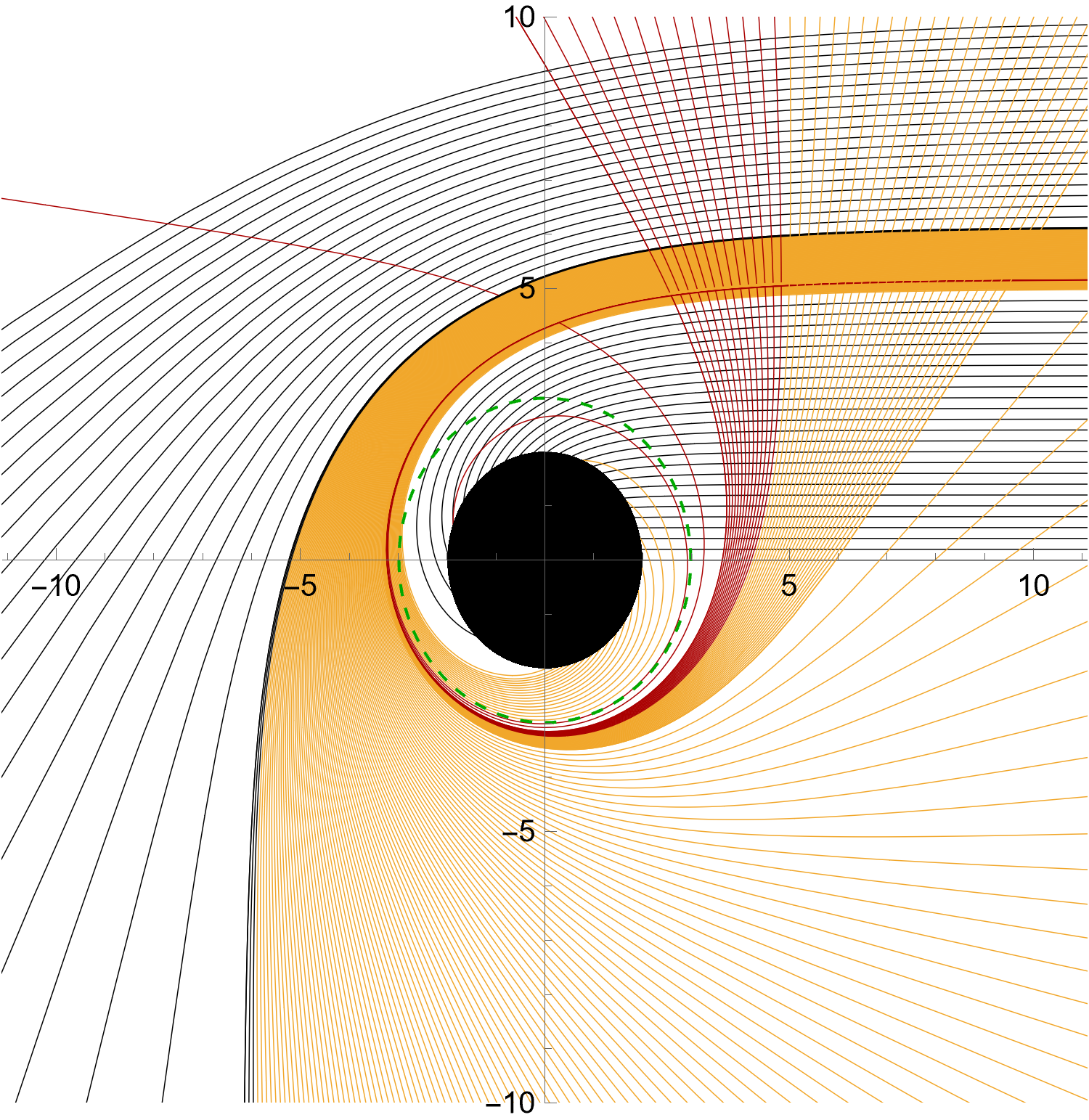}
        \caption{Spacetime traced out by null
geodesics. The plot shows the orbit of light near the
black hole in different colors
correspond to direct orbits $\phi < 3\pi/2$ (black),lensing orbits $3\pi/4 < \phi < 5\pi/4$ (yellow) and photon ring orbits  $\phi > 5\pi/4$ (red) and at the center there is a black disk which is shown the black hole.}
    \label{fig:geodesics}
\end{figure}

The circular null geodesics, as depicted by Fig. \ref{fig:geodesics}, holds the stability condition $V(r)=V'(r)=0$ and $V''(r)>0$ 
For circular photon orbits, the instability is linked to the maximum value of the $V$ as
\begin{equation}
\left.V(r)\right|_{r=r_{p}}=0,\left.\quad V'(r)\right|_{r=r_{p}}=0,
\end{equation}
where, the impact parameter $b \equiv {L}/{E} = {r_{p}}/{\sqrt{f\left(r_{p}\right)}}$ and $r_p=3+\frac{1}{2} \sqrt{36-8 p^2}$ is the radius of the photon sphere: the latter can be computed from the largest root of the relation
\begin{equation} \label{phsp}
\frac{f^{\prime}(r_p)}{f(r_p)}=\frac{2}{r_p},
\end{equation}
Analytically, this is a complicated equation to solve -- numerical methods are employed to obtain the radius $r_p$ of the photon sphere to find that as the value of the confining charge parameter increases, the photon sphere tends to increase as well.

The radius of the shadow of a black hole with respect to a static observer at a position $r_0$ is \cite{EventHorizonTelescope:2020qrl}
\begin{equation}
R_{s}=r_{p} \sqrt{\frac{f\left(r_{0}\right)}{f\left(r_{p}\right)}},
\end{equation}
and for a distant observer ($f(r_{0})=1$), it is
\begin{equation} \label{shadow}
R_{s}^2=\frac{ r_p^{2}}{f(r_p)}.
\end{equation}

Therefore,
\begin{equation}
R_{s}=\frac{\sqrt{2} \left(\sqrt{9-2 p^2}+3\right)}{\sqrt{\frac{p^2+\sqrt{9-2 p^2}-3}{p^2}}}.
\end{equation}

\begin{figure}[htp]
   \centering
    \includegraphics[scale=1]{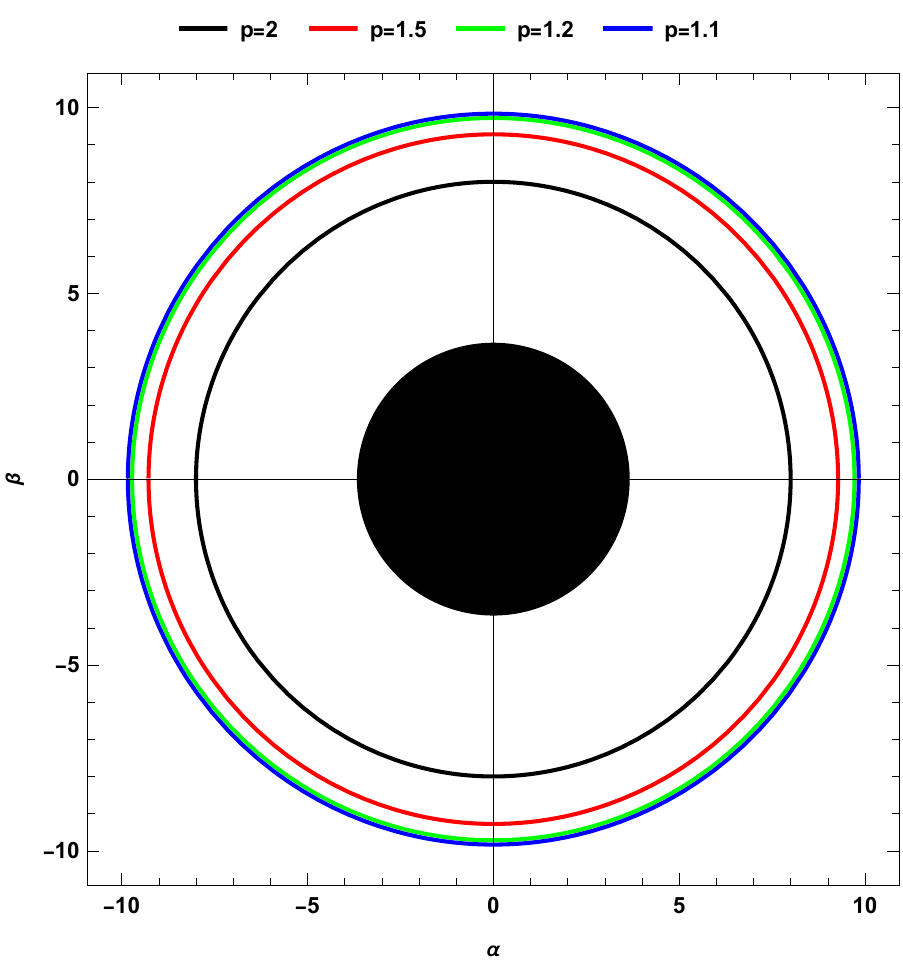}

        \caption{Figure shows the shadow of the black hole with $M=2$ and $\beta=0$ for different values of $p$.}
    \label{fig:shadow1}
\end{figure}

To gather more perspective in this context, Fig. \ref{fig:shadow1} is plotted for different values of $p$. The inverse proportionality between the shadow radius and the magnetic charge can be noticed implying that the greater the magnetic influence, smaller the radius.

\begin{figure}[htp]
   \centering
    \includegraphics[scale=0.8]{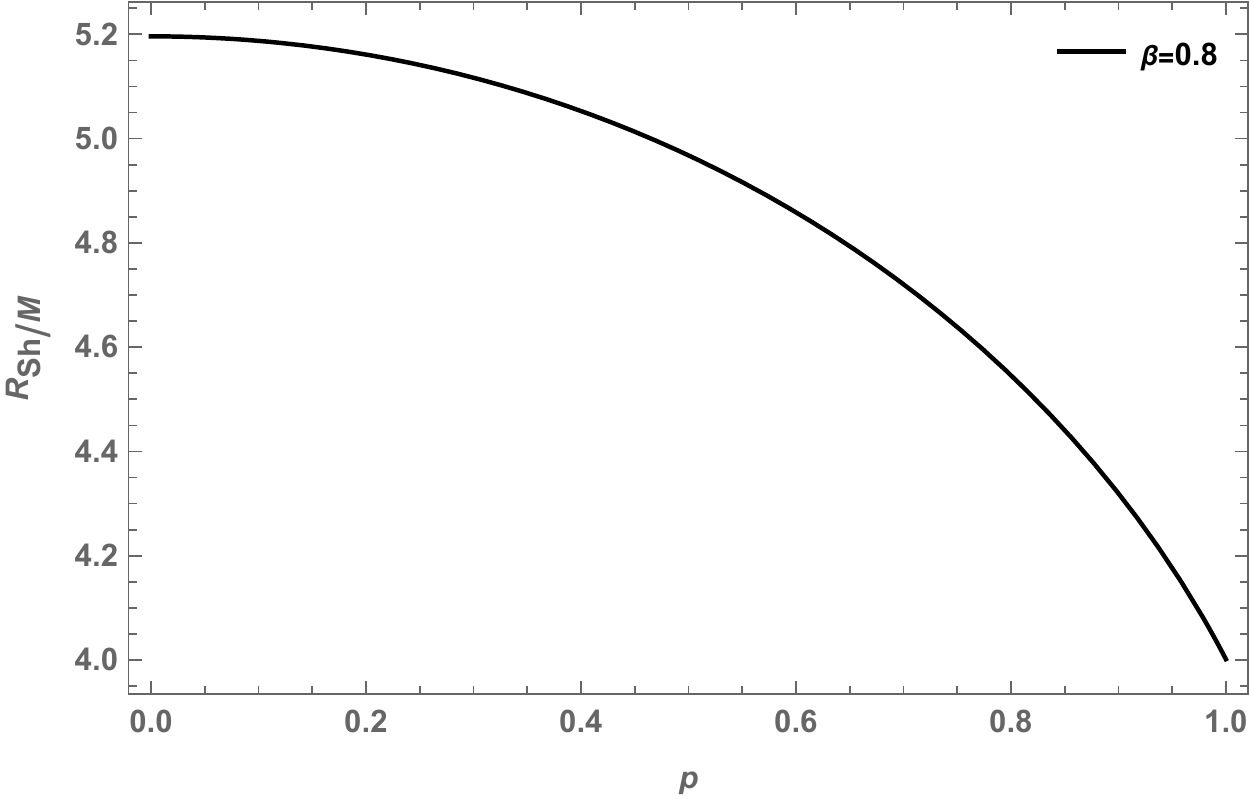}
  
        \caption{The first plot shows the shadow of the black hole with varying $p$ and fixed values of $M=1$ and $\beta=0.8$.}
    \label{fig:shadow2}
\end{figure}

\begin{figure}[htp]
   \centering
    \includegraphics[scale=0.4]{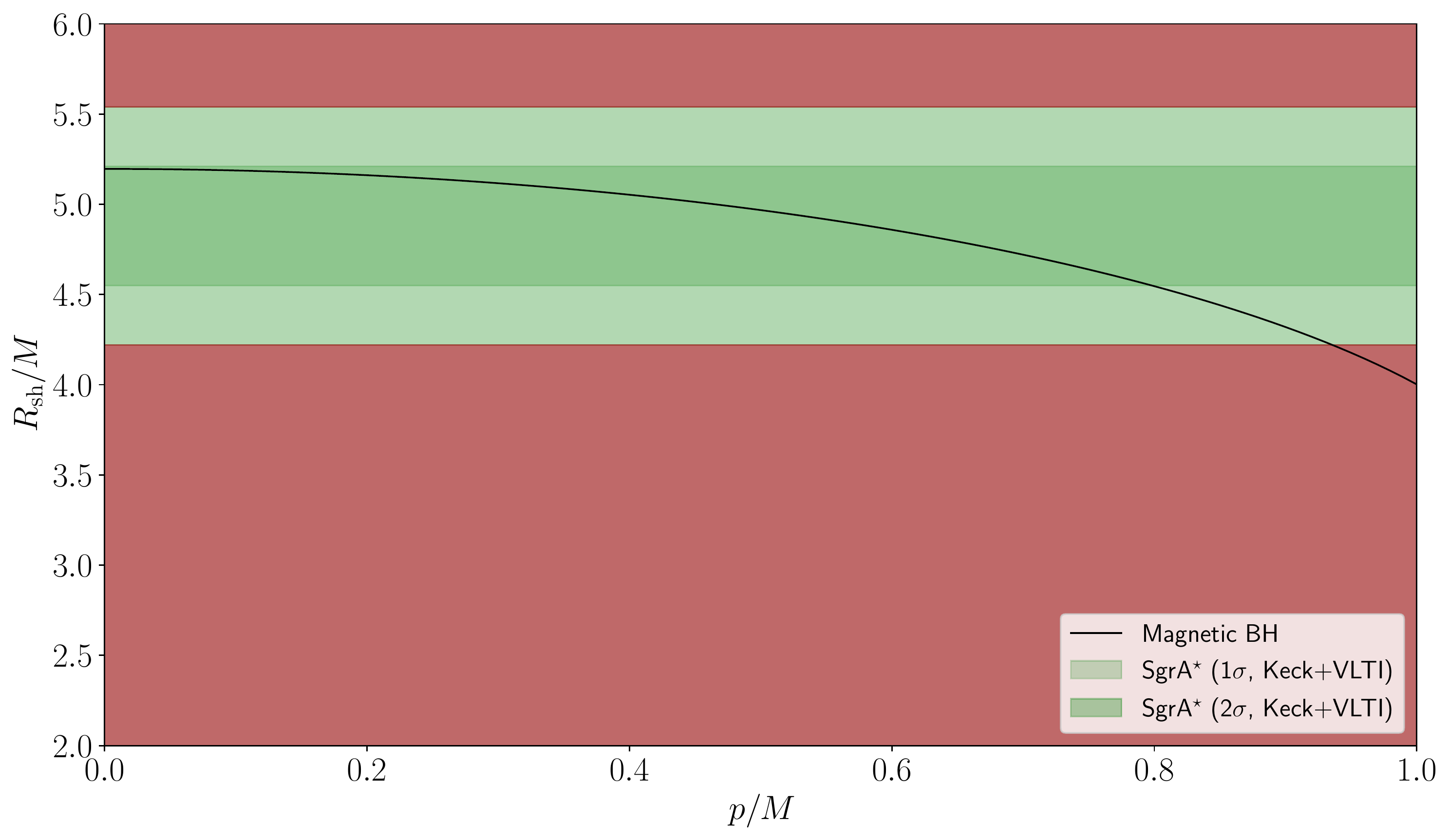}
    \caption{Constraints from EHT  horizon-scale image of SgrA* at
$1\sigma$ and $2\sigma$ \cite{Vagnozzi:2022moj}, after averaging the Keck and VLTI
mass to distance ratio priors for SgrA*   $M=1$, and $\beta=1$ .}
    \label{fig:shadow.eht}
\end{figure}

Fig. \ref{fig:shadow2} is plotted numerically the radius of the black hole ($R_{sh}/M$) shadow with variation of $p$.
The exponential and inverse dependence of $p$ is intriguing, aside from the range of variation that is observed in either of the parameters. Fig. \ref{fig:shadow.eht} shows the upper limit of $p$ from the EHT observations that the $68\%$ confidence
level (C.L.) \cite{Vagnozzi:2022moj} upper limit $p \leq  0.8$ and the $95\%$ C.L.
upper limit $p \leq 0.92$.

Recently, EHT confirmed that the effect of spin for Sgr A* is negligible \cite{EventHorizonTelescope:2022xqj}. Thus, we only study the non-rotating black holes, which is also the same argument that led the authors in Ref. \cite{Vagnozzi:2022moj} to provide a number of constraint plots due various gravity theories and fundamental physics (where $a = 0$). These authors also show that the EHT statement above by providing a visualization that the spin $a$ indeed does not affect the shadow size considerably if $a$ is not so large and the shadow of a rotating black hole is a distorted circle.

\subsection{\label{sec:level5B2} Spherically in-falling accretion}
Following the technique of \cite{Bambi:2013nla}, spherically free-falling accretion is investigated in this section. The accretion disk is dynamic and spherical, unlike the static disc that was inspected before. The number of orbits formalism of \cite{Gralla:2019xty} is implemented again for this dynamic model, except that the crossings are throughout the entire spherical accretion and not at the equatorial plane. The integrated intensity observed at a specific frequency $\nu_{obs}$ expressed as an integral over the null geodesic $\gamma$ to be of the form
\begin{equation}
   I(\nu_{obs},b_\gamma) = \int_\gamma g^3 j(\nu_e) \mathrm{~d}l_{pr},
   \label{eq:bambiI}
\end{equation}
where, $b_\gamma$ is the null geodesic of  the impact parameter, $j$ is the emissivity per unit volume as a function of the emitted frequency, $\mathrm{~d}l_{pr}$ is the infinitesimal (proper) length. and $g$ is the redshift factor altered as
\begin{equation}
    g = \frac{k_\mu u^\mu_o}{k_\mu u^\mu_e}, \; k^\mu = \dot{x}_\mu,
\end{equation}
with $k^\mu$ as the 4-velocity of the photon, $u^\mu_o$ as the 4-velocity of a static observer at infinity, and $u^\mu_e$ as the 4-velocity of the in-falling accretion such that
\begin{equation}
     k_t = \frac{1}{b}, \; k_r = \pm \frac{1}{bf(r)}\sqrt{1 - f(r)\frac{b^2}{r^2}} \quad \textrm{and} \quad u^\mu_e = \left(\frac{1}{f(r)}, -\sqrt{1-f(r)}, 0, 0\right),
\end{equation}
allowing
\begin{equation}
    g = \Big( u_e^t + \frac{k_r}{k_t}u_e^r \Big)^{-1}.
\end{equation}
The proper distance along $\gamma$ can be represented in terms of an affine parameter besides proper time as
\begin{equation}
    \mathrm{~d}l_\gamma = k_\mu u^\mu_e \mathrm{~d}\lambda = \frac{k^t}{g |k_r|}\mathrm{~d}r.
\end{equation}

\begin{figure}[htp]
   \centering
    \includegraphics[scale=0.5]{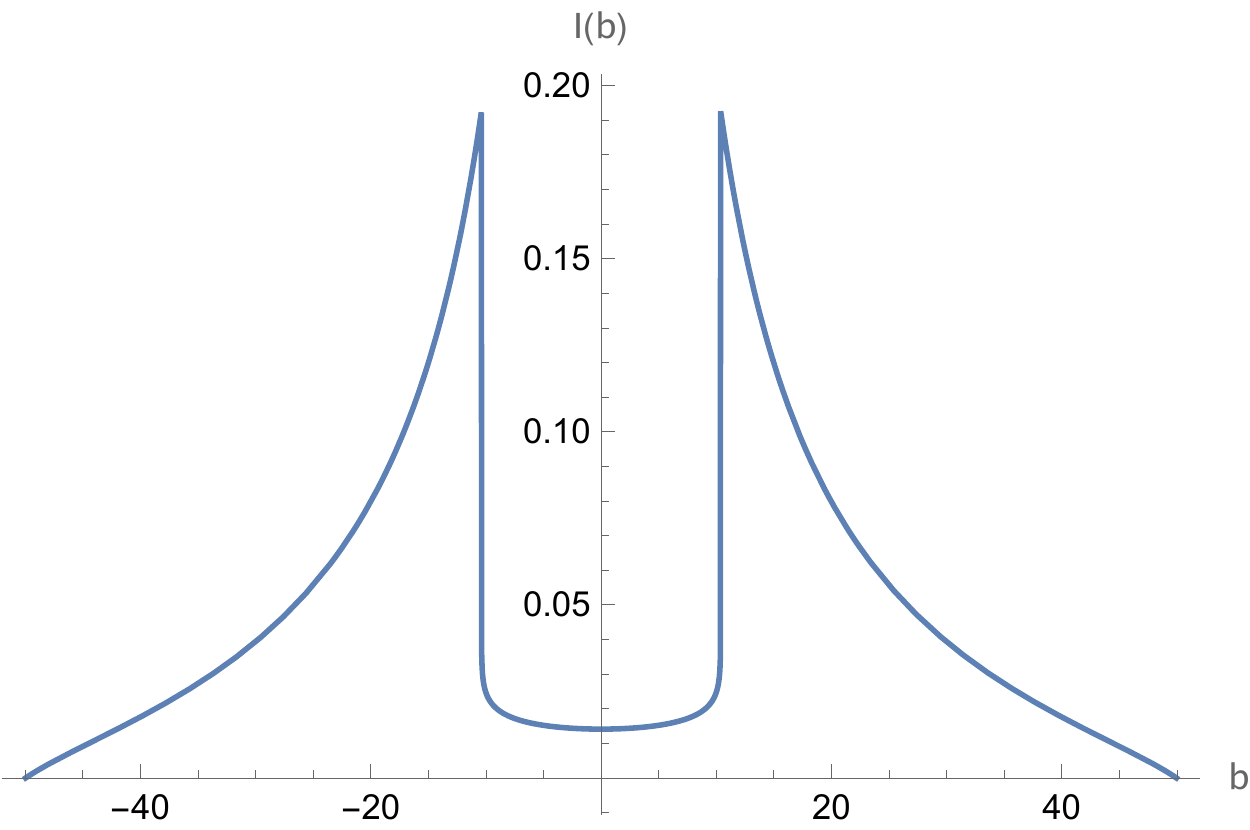}
    \includegraphics[scale=0.5]{
  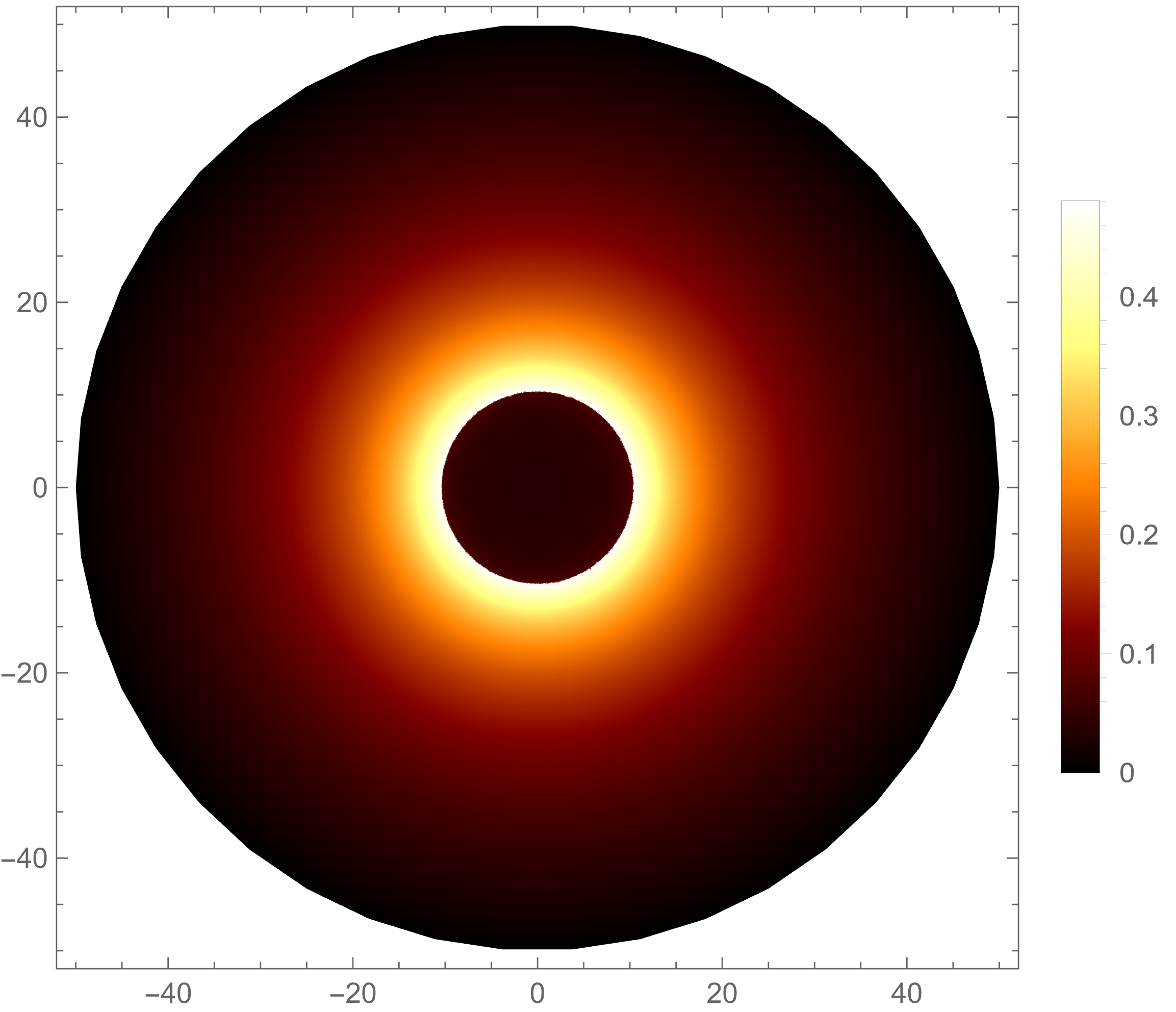}
  
        \caption{Observational appearance of a spherically free-falling accretion emission near a black hole of charge  $M=2$, $p=0.1$ and $\beta=0$.}
    \label{fig:shadow.1}
\end{figure}

\begin{figure}[htp]
   \centering
    \includegraphics[scale=0.6]{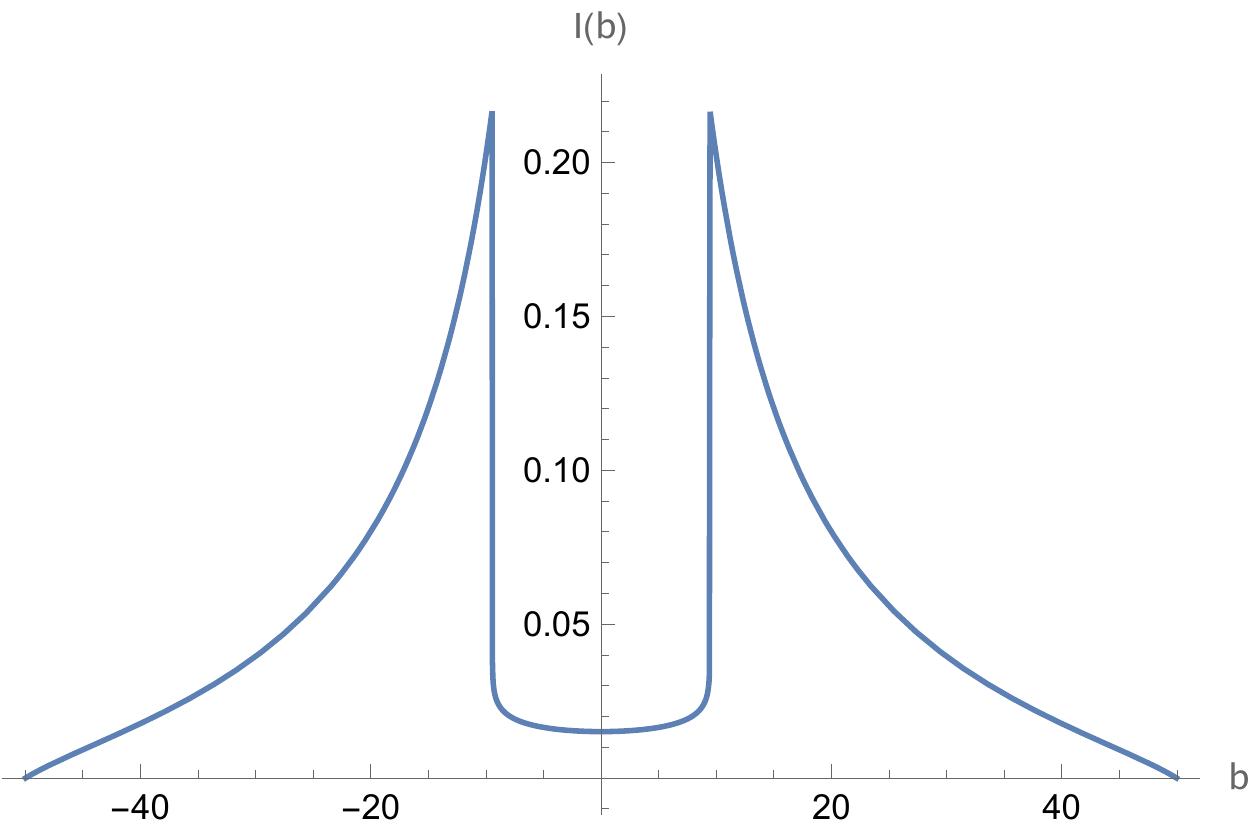}
    \includegraphics[scale=0.6]{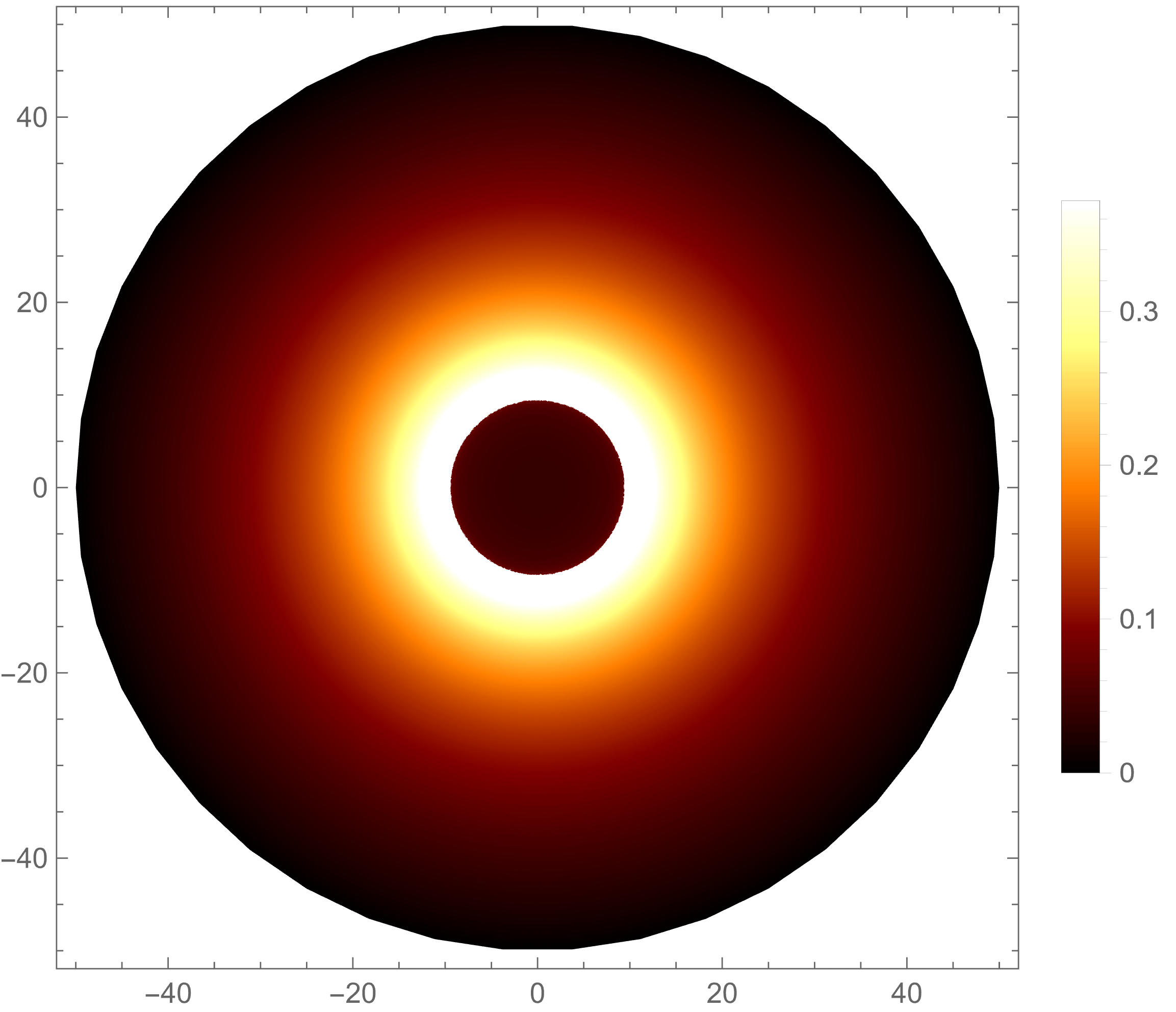}
     \includegraphics[scale=0.8]{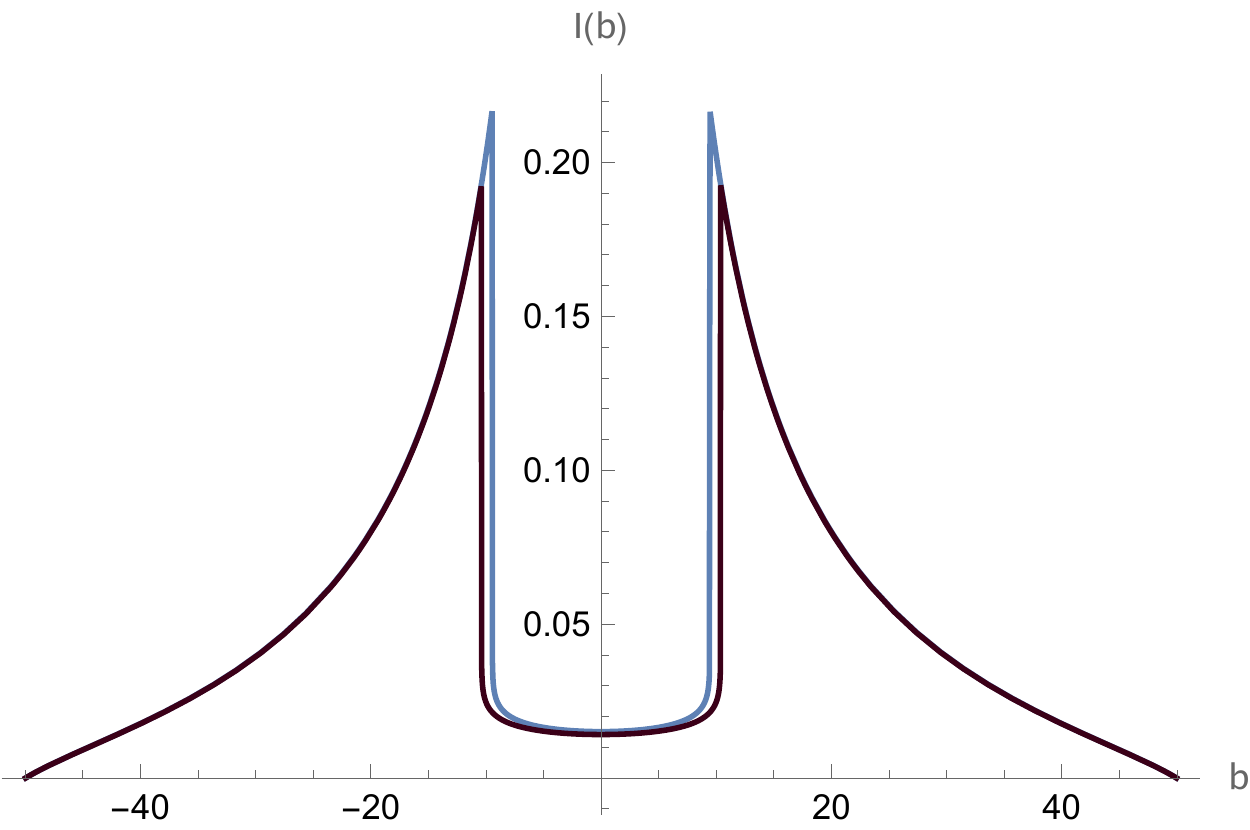}
  
        \caption{Observational appearance of a spherically free-falling accretion emission near a black hole of charge  $M=2$, $p=0.99$ and $\beta=0$. Note that in the last figure, Black line is for ($p=0.1$) and Blue line stands for ($p=0.99$).}
  
    \label{fig:shadow.99}
\end{figure}

For the sake of simplicity, a monochromatic emission with a rest-frame frequency $\nu_*$ and a radial profile ${1}/{r^2}$ is assumed in this model such as
\begin{equation}
    j(\nu_e) \propto \frac{\delta(\nu_e - \nu_*)}{r^2},
\end{equation}
where, $\delta$ is the delta function. Integrating eq.~(\ref{eq:bambiI}) over all frequencies gives the total observed flux to be
\begin{equation}
    F(b_\gamma) \propto \int_\gamma \frac{g^3}{r^2} \frac{k_e^t}{k_e^r} \mathrm{~d}r.
\end{equation}

With the expressions for the flux, we have used Okyay-\"Ovg\"un  \textit{Mathematica} notebook package \cite{Okyay:2021nnh}, (used in \cite{Chakhchi:2022fls,Kuang:2022xjp,Uniyal:2022vdu}) and numerically integrated the flux to see the effects of the charge parameters. See figures \ref{fig:shadow.1}, and  \ref{fig:shadow.99} for examples.

With the help of stereo-graphic projection in the celestial coordinates $X$ and $Y$, the apparent shape of the black hole shadow is plotted in Fig.s \ref{fig:shadow.1} and \ref{fig:shadow.99}. Evidently, the shadow radius increases as value of $p$ increases, proving that the magnetic charge has a strong effect on the black hole shadow size.

Therefore, it is seen that the introduction of a charge term greatly increases the apparent size of the shadow, but decreases the intensity of the incoming light.

\section{Conclusion}
\label{conc}
In this paper, we have examined the weak deflection angle and the shadow of the Reissner-Nordstr\"om black hole under the influence of a magnetic charge for various conditions. We considered higher-order magnetic correction in the Einstein-nonlinear-Maxwell fields to explore the extent of its impact. We defined the function $f(r)$ as the solution of a magnetic black hole.
Then, we calculated the weak deflection angle using the Gauss-Bonnet theorem and we found that the magnetic charge had a direct proportionality to the deflection angle for a given impact parameter. We used the Keeton-Petters formalism to verify our result and extended it to determine three observables, namely, angular position, magnification, and time delay.

In addition to that, we calculated the deflection angle of massive particles using the Jacobi metric in which the velocity of the particles were inversely proportional to the deflection angle and seems to converge into the Schwarzschild case for higher values of the impact parameter. Furthermore, we computed the deflection angle in the presence of plasma and dark matter. The deflection angle was found to be directly dependant on the ratio of electron frequency to photon frequency suggesting that more refraction led to more bending. On the other hand, the deflection decreased with more dark matter activity. Finally, the shadow of a black was studied with respect to the black hole solution, and a special case of a spherically in-falling accretion was probed, studying some parameters of interest. Moreover, we show in Fig. \ref{fig:shadow.eht} that the upper limit of $p$ from the EHT observations that the $68\%$ confidence
level (C.L.) upper limit $p \leq  0.8$, and the $95\%$ C.L.
upper limit  $p \leq 0.92$.

Acknowledgment: A. {\"O}. would like to acknowledge the contribution of the COST Action CA18108 - Quantum gravity phenomenology in the multi-messenger approach (QG-MM).


\vspace{+6pt}
\authorcontributions{Conceptualization, Y.K. and A.Ö.; methodology, A.Ö.; validation, Y.K. and A.Ö.; formal analysis, Y.K.; investigation, Y.K.; resources, Y.K. and A.Ö.;  writing---original draft preparation, Y.K. and A.Ö.; writing---review and editing, Y.K. and A.Ö.; visualization, Y.K. and A.Ö.; supervision, A.Ö. All authors have read and agreed to the published version of the manuscript.}

\funding{This research received no external funding.}

\institutionalreview{Not applicable.}

\informedconsent{Not applicable.}
%

\dataavailability{Not applicable.}

\conflictsofinterest{The authors declare no conflict of interest.} 
\begin{adjustwidth}{-\extralength}{0cm}

\reftitle{References}




\end{adjustwidth}
\end{document}